\documentclass{article}

\usepackage{microtype}
\usepackage{graphicx}
\usepackage{subfig}
\usepackage{booktabs} 
\usepackage{outlines}

\usepackage{hyperref}

\usepackage[accepted]{mlsys2022}

\usepackage{tikz}
\usepackage{scrextend}
\usepackage{amsmath}

\usepackage[english]{babel}
\usepackage{blindtext}
\usepackage{lipsum}
\usepackage{amsthm}
\usepackage[ruled,vlined,linesnumbered,algo2e]{algorithm2e}

\usepackage{filecontents}
\usepackage{xspace}
\usepackage{subfig}
\usepackage{hhline}
\usepackage{multirow}
\usepackage{soul}

\usepackage{array}
\usepackage{textcase}

\usepackage{bbm}
\usepackage{comment}
\usepackage{amsfonts}
\usepackage{makecell}

\usepackage{pifont}
\usepackage{marvosym}

\usepackage[compact]{titlesec}

\newcommand{\cameraold}[1]{{\color{red}{}}}

\newcommand{\jc}[1]{{\color{blue}{\footnotesize [JC: #1]\xspace}}}

\newcommand{\ignore}[1]{{\xspace}}
\newcommand{\REMOVE}[1]{}

\newcommand{\name}{{AccMPEG}\xspace}

\newcounter{packednmbr}

\newenvironment{packeditemize}{\begin{list}{$\bullet$}{\setlength{\itemsep}{0.5pt}\addtolength{\labelwidth}{-4pt}\setlength{\leftmargin}{2ex}\setlength{\listparindent}{\parindent}\setlength{\parsep}{1pt}\setlength{\topsep}{2pt}}}{\end{list}}

\newcommand{\tightcaption}[1]{\vspace{-0.1cm}\caption{{\normalfont{\textit{{#1}}}}}\vspace{-0.0cm}}
\newcommand{\tightsection}[1]{\vspace{-0.17cm}\section{#1}\vspace{-0.25cm}}

\newcommand{\tightsubsection}[1]{\vspace{-0.2cm}\subsection{#1}\vspace{-0.2cm}}

\newcommand{\eg}{{\it e.g.,}\xspace}
\newcommand{\ie}{{\it i.e.,}\xspace}

\newcommand{\mypara}[1]{\noindent{\bf {#1}:}~}

\begin{document}

\date{}

\twocolumn[
\mlsystitle{\name: Optimizing Video Encoding for Video Analytics}


\mlsyssetsymbol{equal}{*}

\begin{mlsysauthorlist}
\mlsysauthor{Kuntai Du}{uchi}
\mlsysauthor{Qizheng Zhang}{uchi}
\mlsysauthor{Anton Arapin}{uchi}
\mlsysauthor{Haodong Wang}{uchi}
\mlsysauthor{Zhengxu Xia}{uchi}
\mlsysauthor{Junchen Jiang}{uchi}
\end{mlsysauthorlist}

\mlsysaffiliation{uchi}{University of Chicago}
\mlsyscorrespondingauthor{Kuntai Du}{kuntai@uchicago.edu}

\mlsyskeywords{Video Analytics, Video Streaming, Video Compression}


\begin{abstract}

With more videos being recorded by edge sensors (cameras) and analyzed by computer-vision deep neural nets (DNNs), a new breed of video streaming systems has emerged, with the goal to compress and stream videos to remote servers in real time while preserving enough information to allow highly accurate inference by the server-side DNNs. 
An ideal design of the video streaming system should simultaneously meet three key requirements:
(1) low latency of encoding and streaming, 
(2) high accuracy of server-side DNNs, and 
(3) low compute overheads on the camera.
Unfortunately, despite many recent efforts, such video streaming system has hitherto been elusive, especially when serving advanced vision tasks such as object detection or semantic segmentation.

This paper presents \name, a new video encoding and streaming system that meets all the three requirements.
The key is to learn how much the encoding quality at each (16x16) macroblock can influence the server-side DNN accuracy, which we call {\em accuracy gradient}.
Our insight is that these macroblock-level accuracy gradient can be inferred with sufficient precision by feeding the video frames through a cheap model.
\name provides a suite of techniques that, given a new server-side DNN, can quickly create a cheap model to infer the accuracy gradient on any new frame in near realtime.
Our extensive evaluation of \name on two types of edge devices (one Intel Xeon Silver 4100 CPU or NVIDIA Jetson Nano) and three vision tasks (six recent pre-trained DNNs) shows that \name (with the same camera-side compute resources) can reduce the end-to-end inference delay by 10-43\% without hurting accuracy compared to the state-of-the-art baselines. 

\end{abstract}
]

\printAffiliationsAndNotice{}


\newcommand{\HighQuality}{\ensuremath{HQ}\xspace}
\newcommand{\LowQuality}{\ensuremath{LQ}\xspace}
\newcommand{\MaxHighQualityFrac}{\ensuremath{c}\xspace}
\newcommand{\OriginWidth}{\ensuremath{W}\xspace}
\newcommand{\OriginHeight}{\ensuremath{H}\xspace}
\newcommand{\GridWidth}{\ensuremath{w}\xspace}
\newcommand{\GridHeight}{\ensuremath{h}\xspace}
\newcommand{\Mask}{\ensuremath{\textrm{\bf M}}\xspace}
\newcommand{\HighImage}{\ensuremath{\textrm{\bf H}}\xspace}
\newcommand{\LowImage}{\ensuremath{\textrm{\bf L}}\xspace}
\newcommand{\RawImage}{\ensuremath{\textrm{\bf R}}\xspace}
\newcommand{\AllOnes}{\ensuremath{\textrm{\bf 1}}\xspace}
\newcommand{\DNN}{\ensuremath{D}\xspace}
\newcommand{\AccFunc}{\ensuremath{Acc}\xspace}
\newcommand{\AccuracyFunc}{\ensuremath{Acc}\xspace}
\newcommand{\AccuracyGradient}{\ensuremath{AccGrad}\xspace}
\newcommand{\AccuracyGradients}{\ensuremath{AccGrads}\xspace}
\newcommand{\AccuracyGradientModel}{\ensuremath{AccModel}\xspace}
\newcommand{\SimFunc}{\ensuremath{Acc}\xspace}
\newcommand{\Image}{\ensuremath{\textrm{\bf X}}\xspace}
\newcommand{\AgThresh}{\ensuremath{\alpha}\xspace}
\newcommand{\AreaThresh}{\ensuremath{\beta}\xspace}
\newcommand{\PadThresh}{\ensuremath{\gamma}\xspace}
\newcommand{\ag}{\ensuremath{AccGrad}\xspace}
\newcommand{\ags}{\ensuremath{AccGrad}\xspace}
\newcommand{\Ag}{\ensuremath{AccGrad}\xspace}
\newcommand{\Ags}{\ensuremath{AccGrad}\xspace}
\newcommand{\Model}{\ensuremath{M}\xspace}
\newcommand{\MaskGen}{\ensuremath{F}\xspace}
\newcommand{\Block}{\ensuremath{B}\xspace}
\newcommand{\Pixel}{\ensuremath{i}\xspace}
\newcommand{\Edit}{}

\newcommand{\red}[1]{{#1}}
\newcommand\myeq{\stackrel{\mathclap{\normalfont\mbox{def}}}{=}}



\tightsection{Introduction}

Empowered by modern computer vision, video analytics applications running on edge/mobile devices are poised to transform businesses (retail, industrial logistics, home assistance, etc), and public policies (traffic management, urban planning, etc)~\cite{trafficvision, traffictechnologytoday, goodvision,intuvisiontech, vision-zero, msr}.
These emerging video applications use deep neural networks (DNNs) to analyze massive videos from edge video sensors, resulting in an explosive growth of video data~\cite{infowatch12,infowatch16} that serve {\em analytical} purposes rather than being watched by human users for entertainment~\cite{infowatch16,VideoAnalyticsMarket,slate-video-news}.

A key component of these video analytics applications is an efficient {\em video compression}\footnote{We use video compression and encoding interchangeably.} algorithm, which compresses videos in realtime while preserving enough information for accurate inference by the {\em final DNN} running on a remote (cloud) server~\cite{vcm-news}. 
An ideal video compression algorithm should meet three requirements that are key to edge video analytics applications:
(1) high inference accuracy by the final DNN,
(2) low end-to-end delay of encoding and streaming the video to the analytical server, and 
(3) low compute overhead on the camera.
There have recently been many proposals of such analytics-oriented video compression algorithms; for instance, they leverage the {\em spatial heterogeneity} of the final DNNs~\cite{dds,eaar}, such as object detection and semantic segmentation: 
only in a small fraction of regions (\eg which contain important details), lowering encoding quality tends to lower inference accuracy and renders it useless.

Unfortunately, none of the existing solutions can simultaneously meet all three requirements, especially when serving advanced tasks such as object detection or semantic segmentation.
For instance, using camera-side heuristics to filter out (or lower the encoding quality of) unuseful pixels quality is sensible~\cite{vigil,reducto,filterforward}, but existing designs cannot precisely lower the encoding quality of unuseful pixels without affecting useful ones, unless the heuristics themselves are almost as compute-intensive as the final DNNs.
In response, some proposals achieve high accuracy and low camera-side overhead by sending the content to the server-side DNN first to extract feedback~\cite{dds,eaar,elf} from the server, based on which the camera can then encode the video near-optimally or run local inference (\eg object tracking),
causing high end-to-end delay.
Other solutions extract and encode feature maps on the camera~\cite{duan2020video,xia2020emerging,cracking,neurosurgeon,matsubara2019distilled}, and they work well for classification models but not for advanced DNNs (object detection and segmentation models), whose feature maps are orders of magnitude larger than those of classification DNNs.

This paper presents a new video streaming system called \name that meets the three aforementioned requirements. 
At a high level, \name runs a cheap {\em quality selector} logic (a shallow neural net, MobileNet-SSD~\cite{ssd,mobilenetv2}) that determines a near-optimal encoding scheme for any frame---the encoding quality at each (16x16) macroblock, and encodes the frames using popular video codecs like H.26x with region-of-interest (RoI) encoding.
The insight underpinning the cheap quality selector is that inferring the influence of the encoding quality of each macroblock on the final DNN accuracy, which we call {\em accuracy gradient} or \AccuracyGradient, is much simpler than semantic segmentation (a common vision task) for multiple reasons (\S\ref{subsec:accuracy-gradient}): for instance, assigning high or low quality to macroblocks of a 720p frame (1280x720) is equivalent to assigning binary labels on an 80x45 image (a 720p frame has 80x45 macroblocks), which can be much simpler than most modern vision tasks.



\name's quality selection also strikes a favorable accuracy-delay balance.
Prior techniques assign encoding quality at coarser granularities, such as encoding entire bounding boxes in high quality and entire background in low quality. 
In contrast, \name's macroblock-level quality selection could outperform them by encoding some background macroblocks in high quality (\eg to provide necessary context that improves accuracy) and some inside the bounding boxes in low quality (\eg if encoding other macroblocks of a large object in high quality is sufficient to achieve high accuracy).

Finally, one must be able to quickly customize the quality selector based on the need of any new final DNN. 
Traditionally, training such a  quality selector 
requires running though an entire pipeline (quality selection, encoding, and inference) over many training images. 
In contrast, \name \ {\em decouples} the final DNN from the training of the quality selector (\S\ref{sec:training}).
To this end, we directly derive the \AccuracyGradient per macroblock by treating the final DNN as a differentiable blackbox, and 
then train the quality selector as a standalone (low-dimensional) segmentation model to infer these \AccuracyGradient, which also reduces the number of training iterations and training images.

\begin{figure}
    \centering
    \includegraphics[width=0.45\columnwidth]{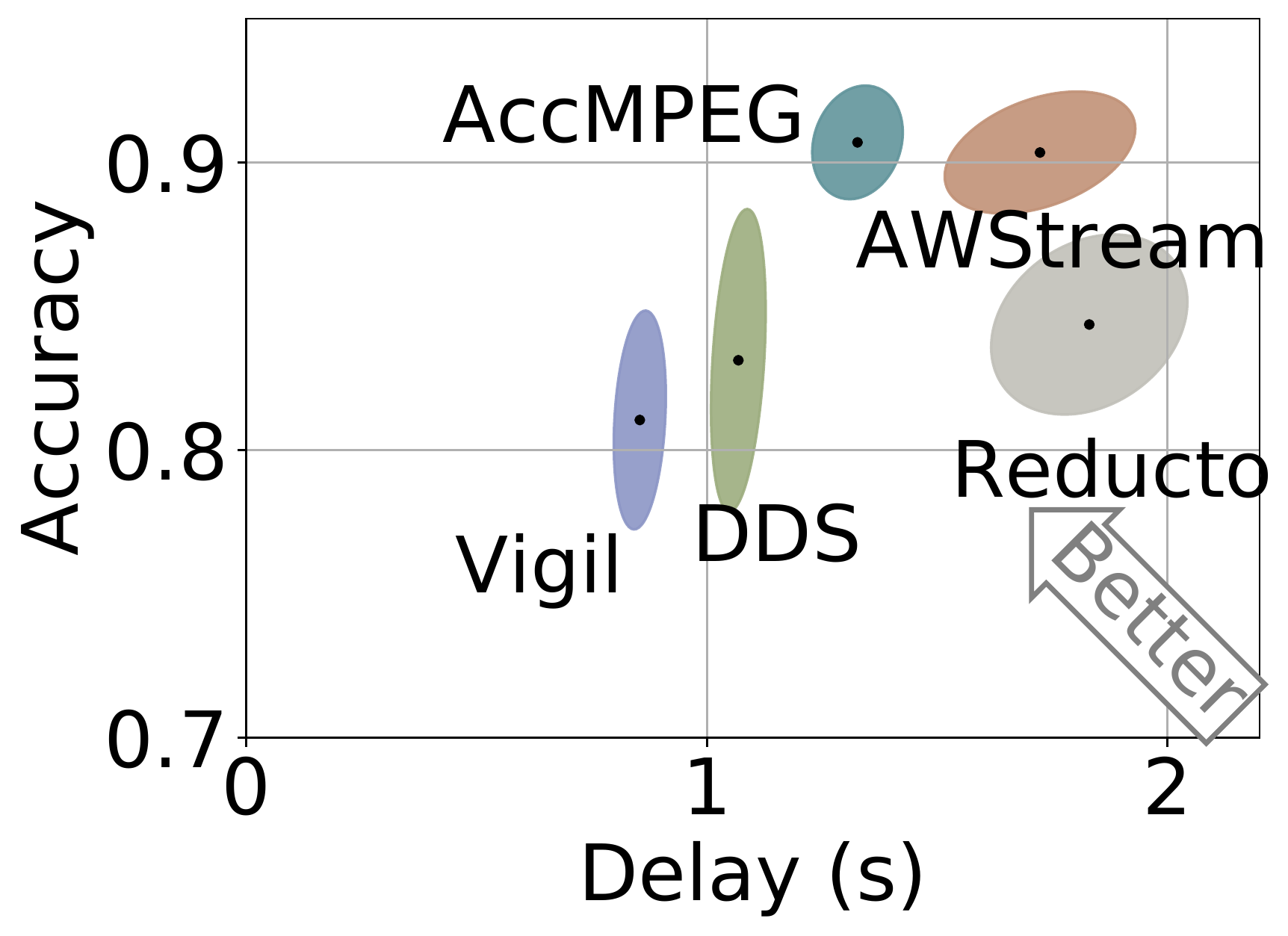}
    \includegraphics[width=0.45\columnwidth]{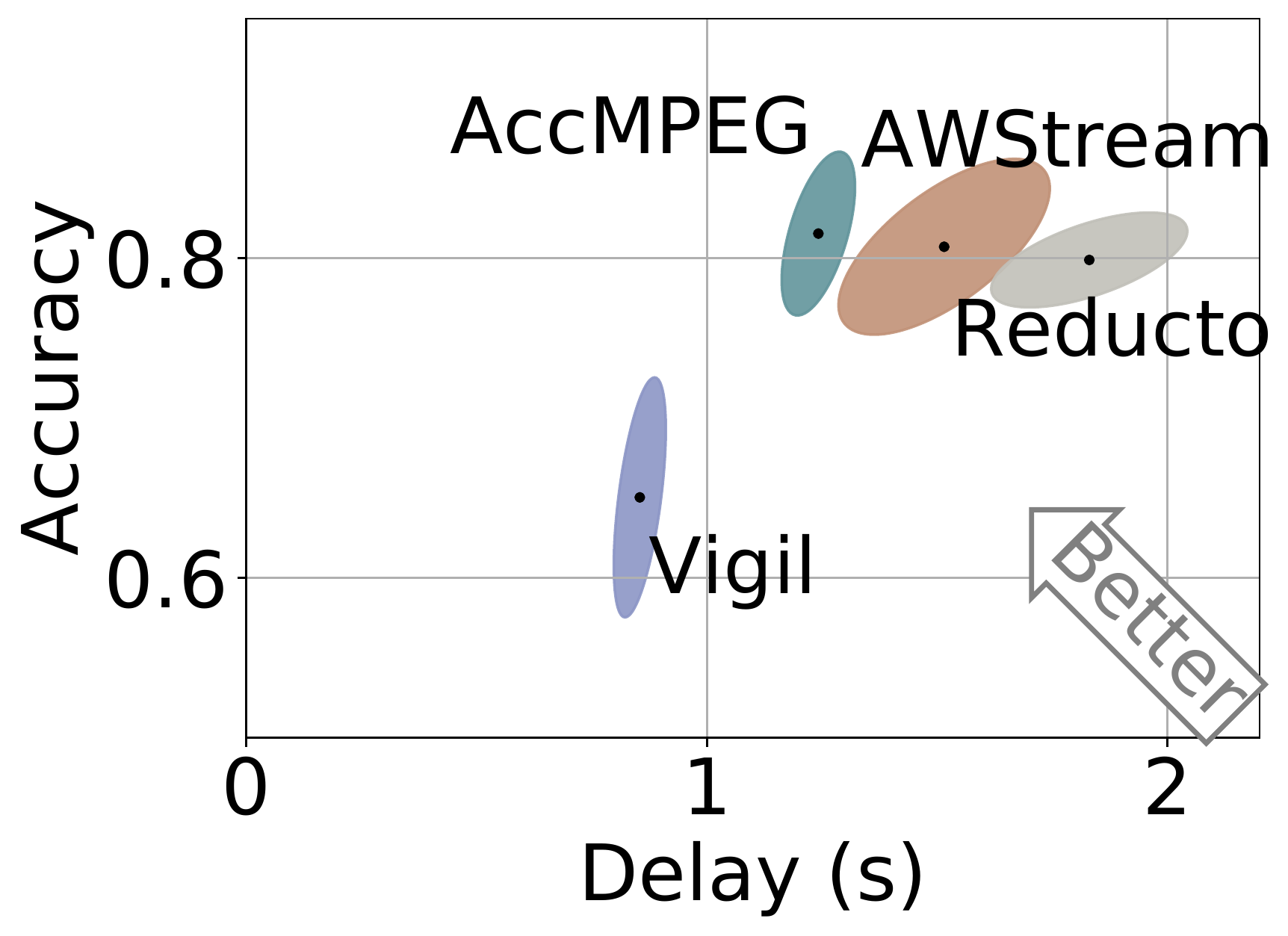}
    \tightcaption{Example results of the accuracy-delay tradeoffs of \name and baselines. \name achieves 10-30\% smaller end-to-end delay  without sacrificing accuracy, or 1-5\% higher inference accuracy than state-of-the-art solutions.}
    \label{fig:result-overview}
\end{figure}
\begin{figure*}[t]
    \centering
    \subfloat[][{FasterRCNN (Highway)}]
    {
        \includegraphics[width=0.22\textwidth]{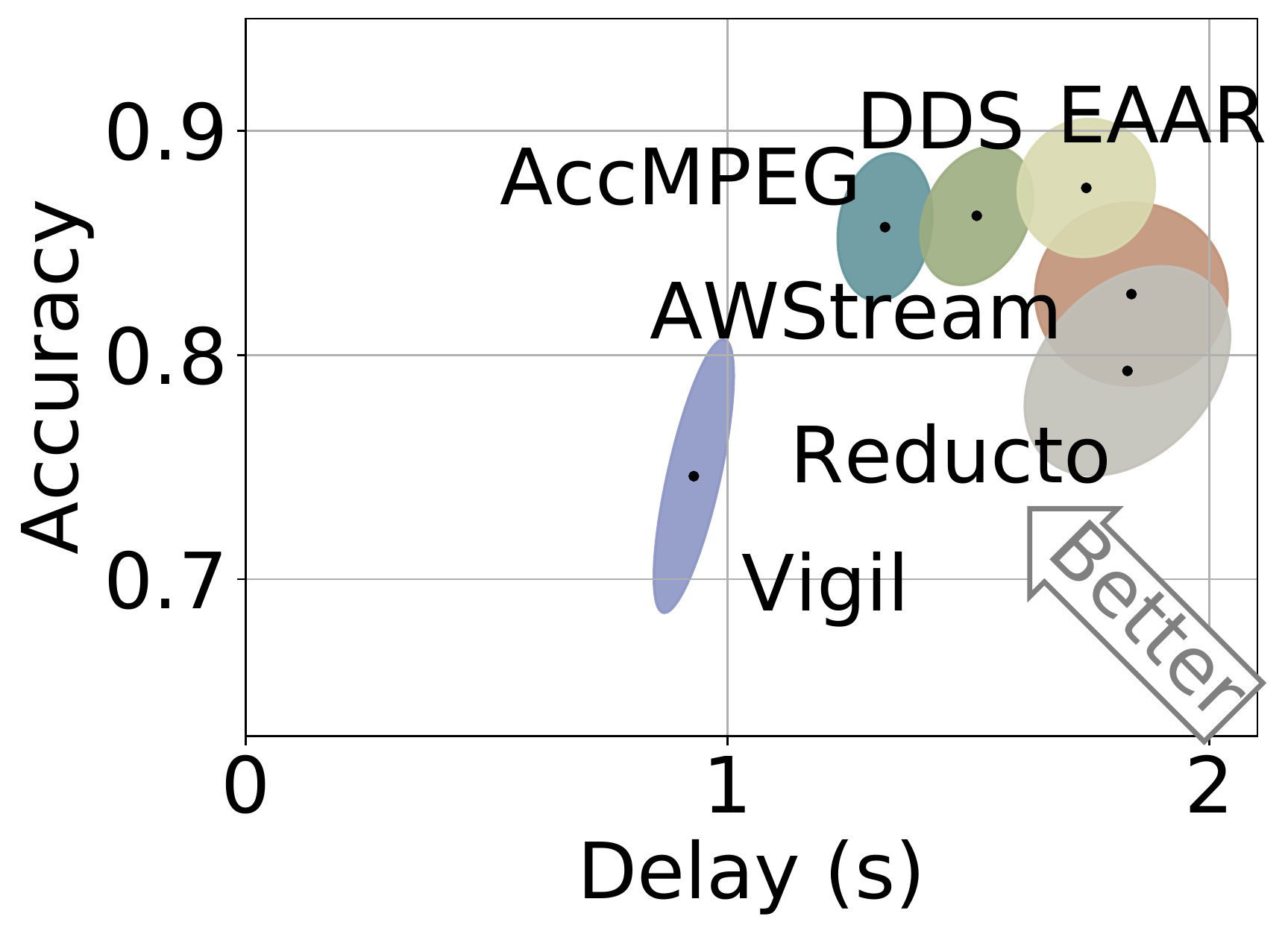}
        \label{subfig:1}
    }
    \subfloat[][{YoLo (Highway)}]
    {
        \includegraphics[width=0.22\textwidth]{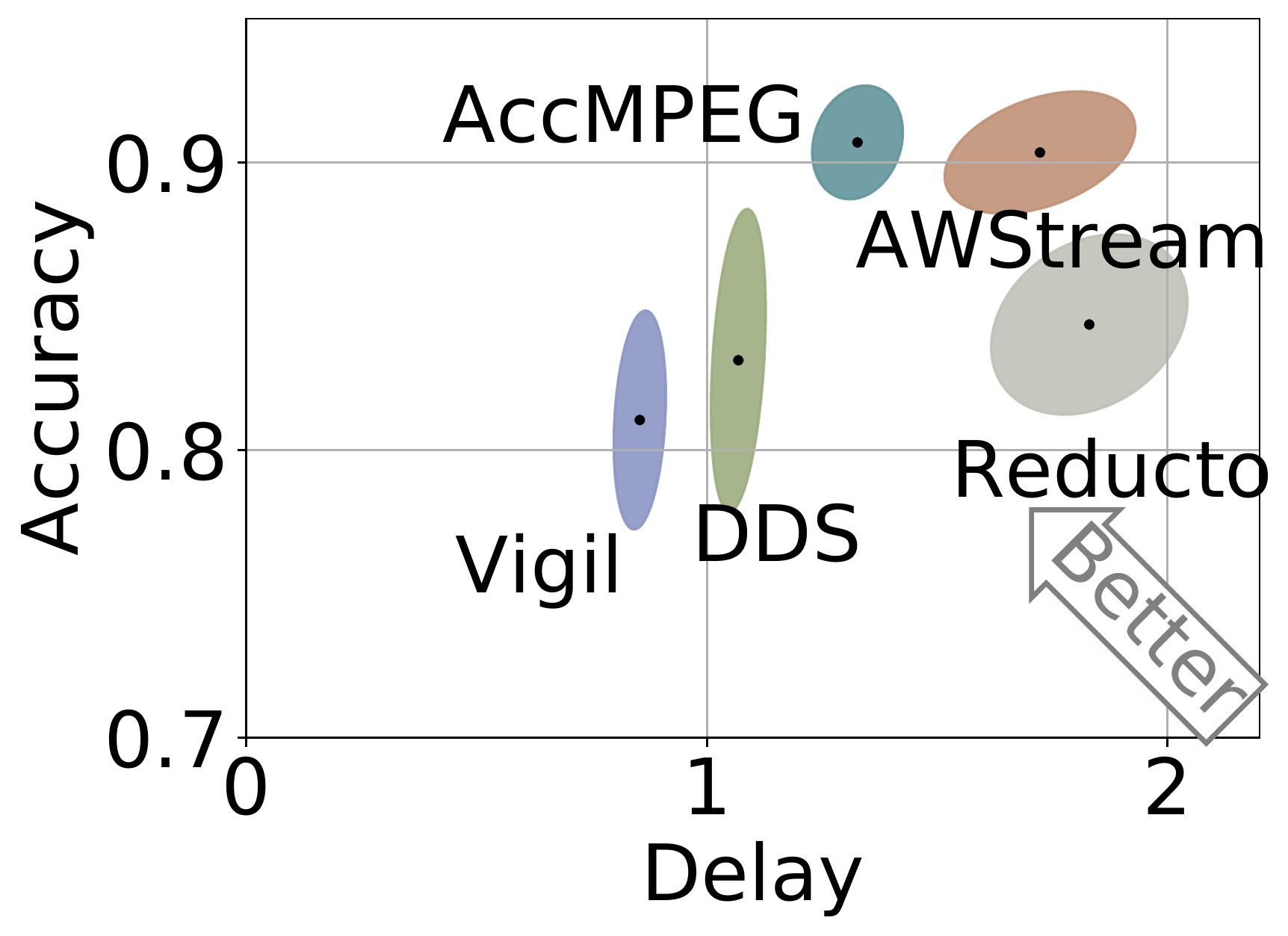}
        \label{subfig:2}
    }
    \subfloat[][{EfficientDet (Highway)}]
    {
        \includegraphics[width=0.22\textwidth]{figs/detection_dashcam_EfficientDet.pdf}
        \label{subfig:1}
    }
    \subfloat[][{Keypoint-R50 (Surf)}]
    {
        \includegraphics[width=0.22\textwidth]{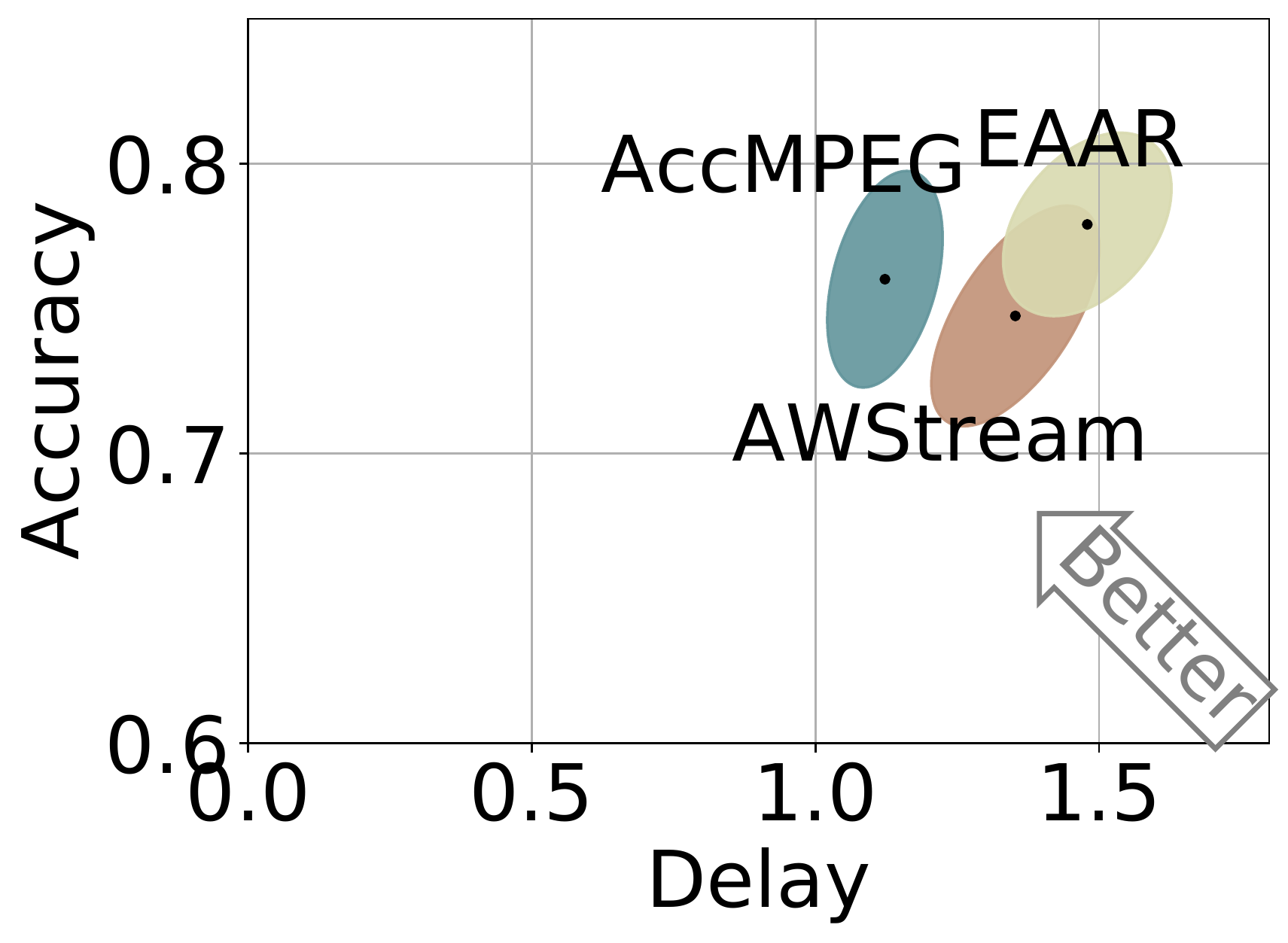}
        \label{subfig:2}
    }
    \\
    \subfloat[][FasterRCNN (Driving)]
    {
        \includegraphics[width=0.22\textwidth]{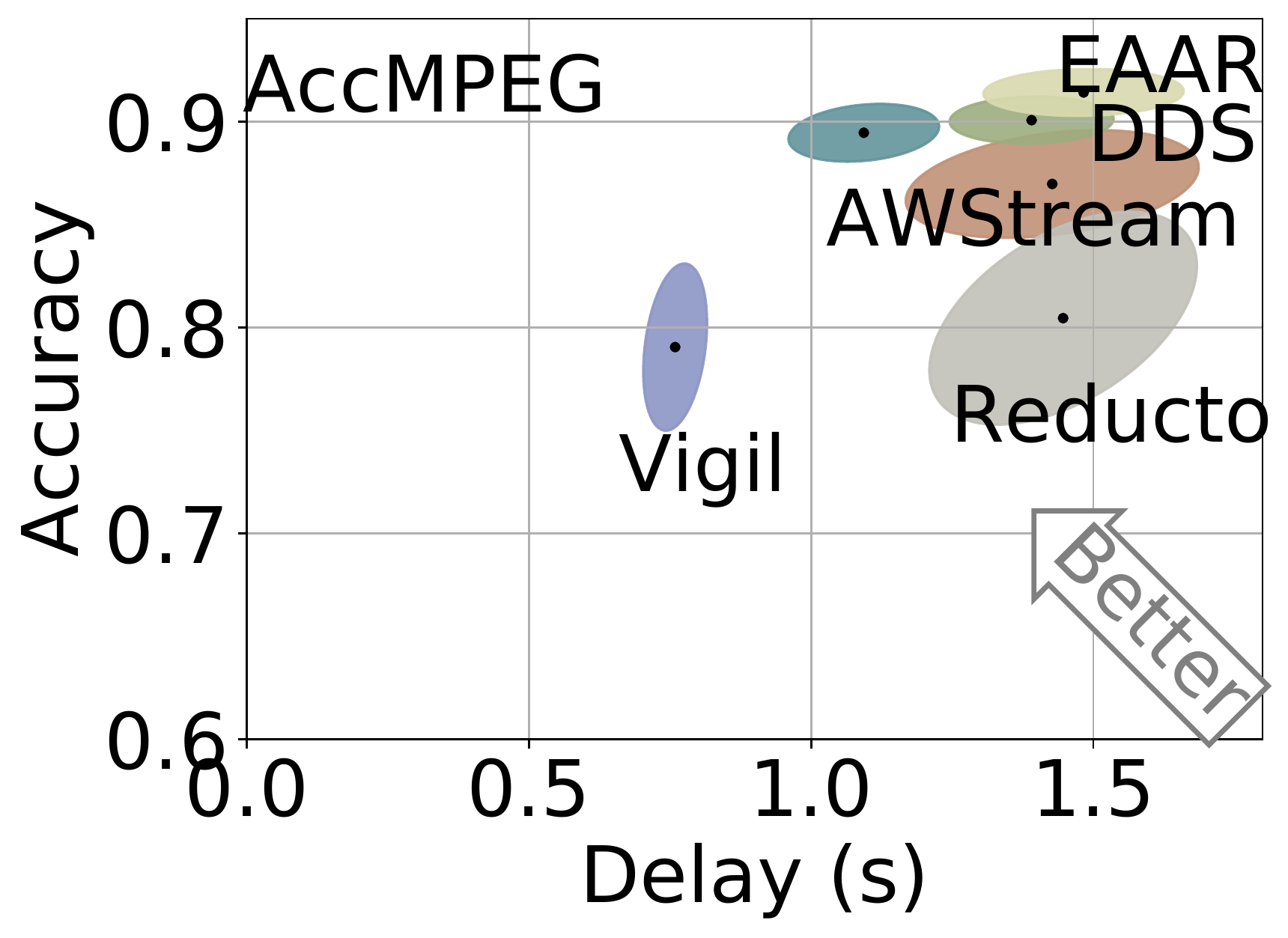}
        \label{subfig:e}
    }
    \subfloat[][YoLo (Driving)]
    {
        \includegraphics[width=0.22\textwidth]{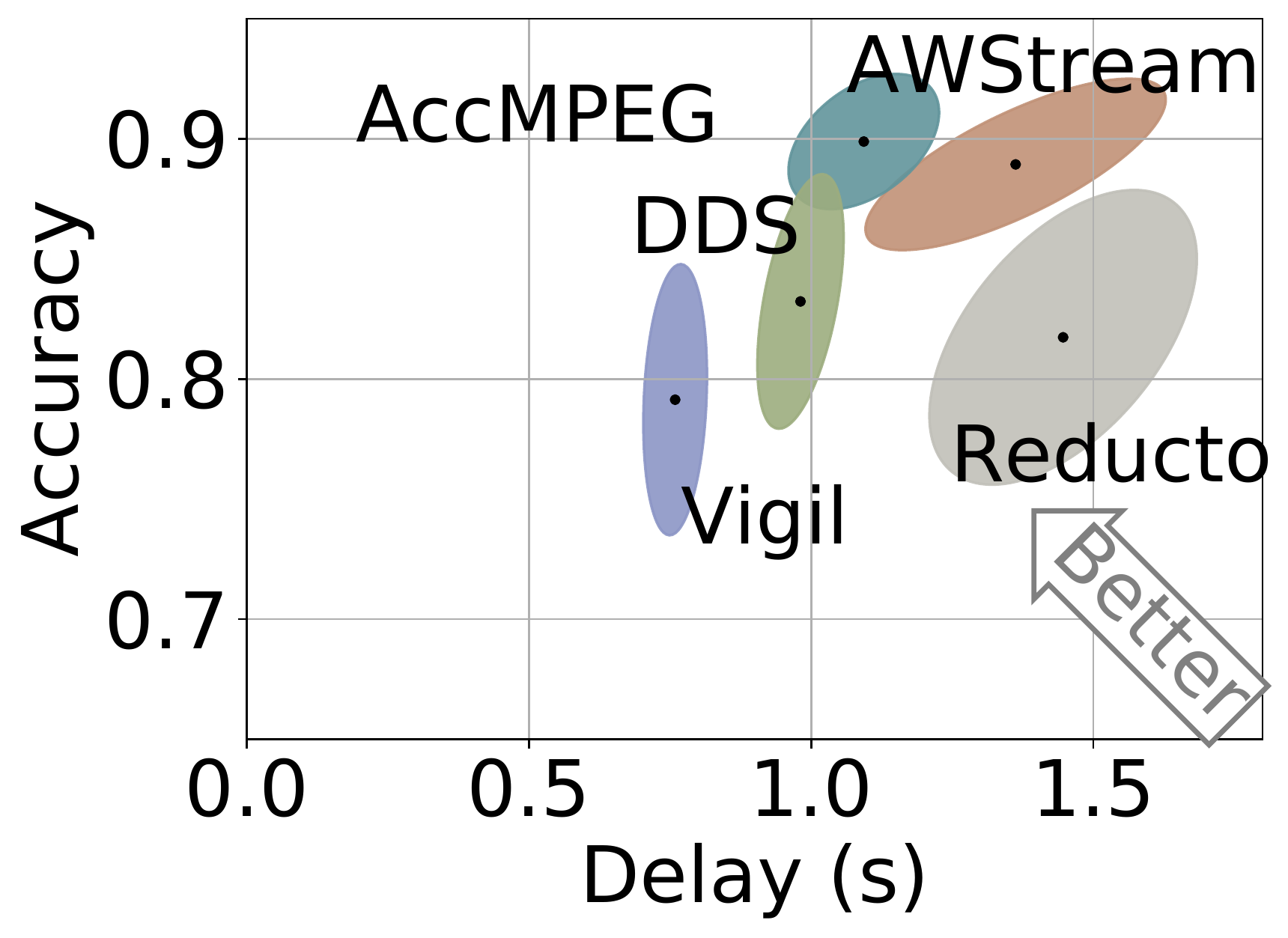}
        \label{subfig:1}
    }
    \subfloat[][EfficientDet (Driving)]
    {
        \includegraphics[width=0.22\textwidth]{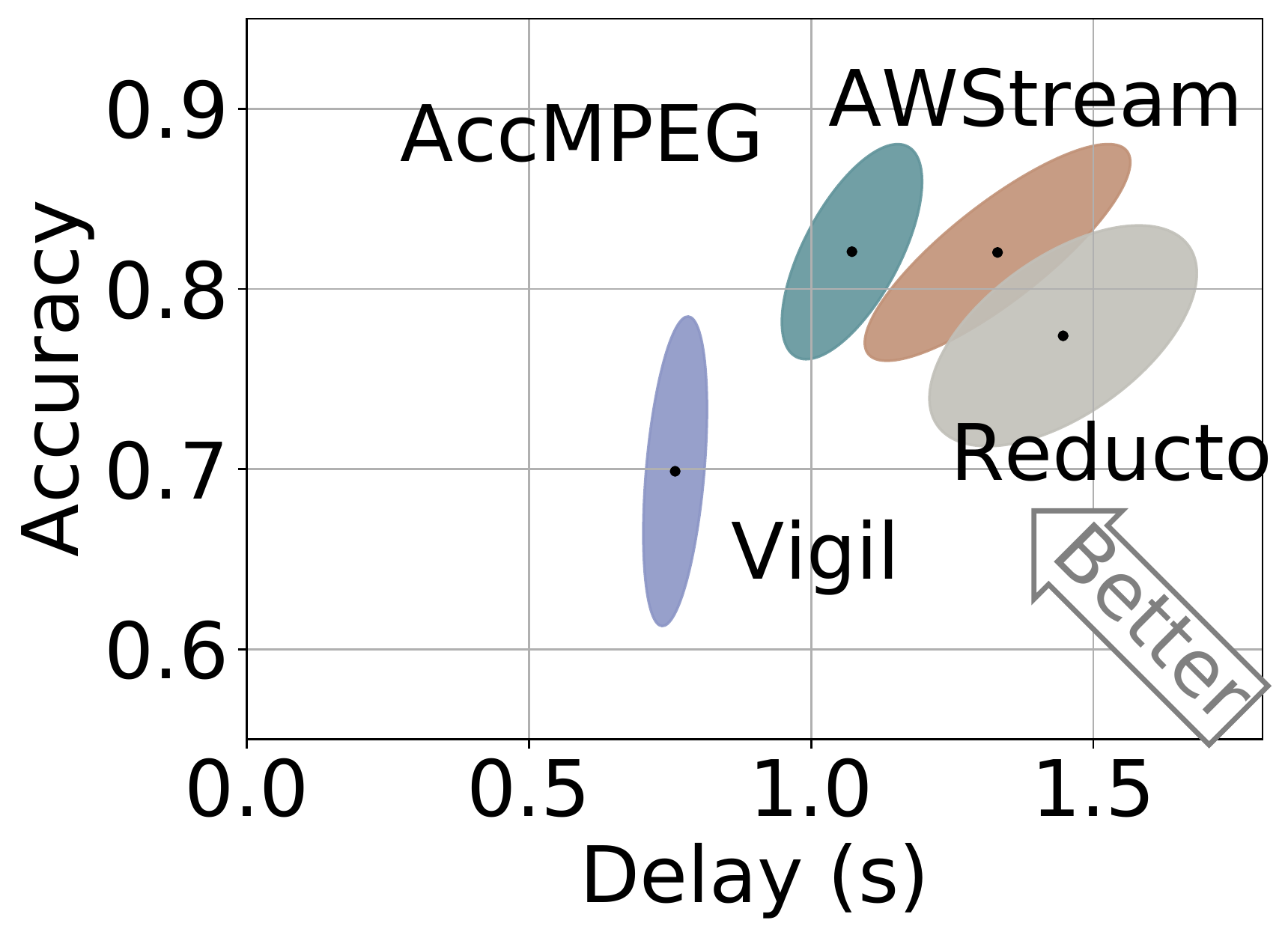}
        \label{subfig:1}
    }
    \subfloat[][Keypoint-X101 (Surf)]
    {
        \includegraphics[width=0.22\textwidth]{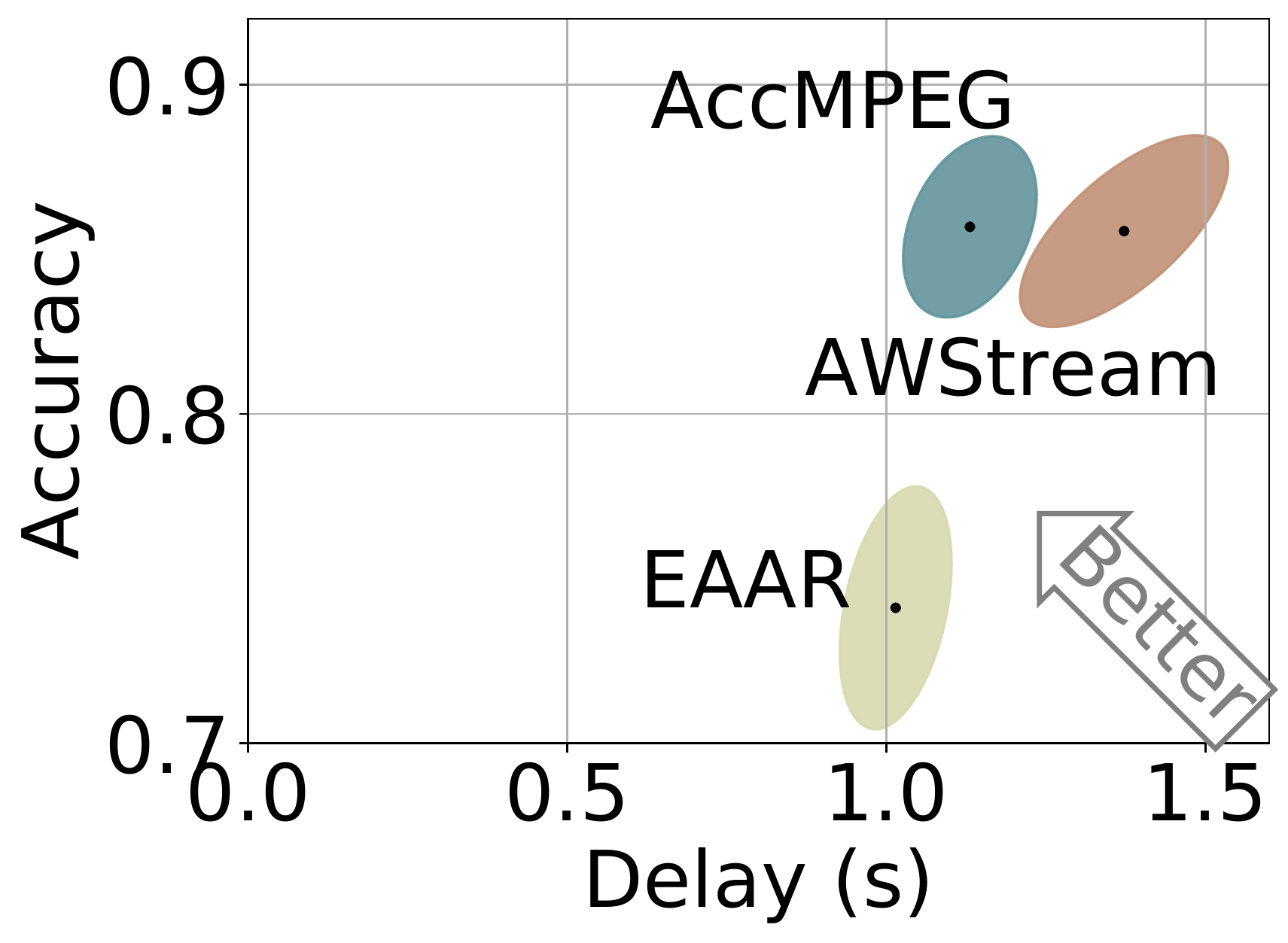}
        \label{subfig:1}
    }
    \tightcaption{The delay vs. inference accuracy of \name and several baselines on various video datasets (in parentheses) and different DNN models (the three object detection models use different backbones). 
    \name achieves high accuracy with 10-43\% delay reduction on object detection and 17\% on keypoint detection.
    Ellipses show the 1-$\sigma$ range of results.}
    \label{fig:eval-overall}
\end{figure*}

Using videos of three different genres, three typical vision tasks (object detection, semantic segmentation, and keypoint detection), and five recent off-the-shelf vision DNNs, we show that on two types of edge devices (one CPU or a Jetson Nano) 
\name can reduce the inference delay by 10-43\% without hurting accuracy compared to various state-of-the-art baselines.
Figure~\ref{fig:result-overview} shows an example improvement of \name's accuracy-delay tradeoffs over the baselines (See Figure~\ref{fig:eval-overall} for the complete results).
\name's encoding speed (30fps with one Intel Xeon Silver 4100 CPU) is only marginally slower than basic video encoding and is much faster than the baselines that achieve similar compression efficiency.
Moreover, an \name encoder for a new DNN can be created within only 8 minutes.


Admittedly, not all techniques of \name are exactly new: 
it uses standard video codec libraries~\cite{h264,h265} that support RoI encoding, 
and many proposals in this space (\eg~\cite{wang2017residual,mnih2014recurrent}) use the the spatially uneven distribution of DNN attention.
Nonetheless, \name strikes a unique balance among encoding/streaming delay, inference accuracy, and low compute overhead in advanced vision tasks (\eg object detection, segmentation). 
Our contribution is two-fold: 
\begin{packeditemize}
\item A cheap quality selector that infers accuracy gradient  (how sensitive a DNN's output is to the encoding quality of each macroblock) on each frame and selects encoding quality per macroblock, with a compute overhead only on par with video encoding. 
\item Fast training (within several minutes) of the quality selector for any given final DNN, which degrades gracefully when the final DNN changes.
\end{packeditemize}

\tightsection{Motivation}
\label{sec:motivation}

We begin by motivating the three performance requirements that drive our design (low accuracy, low encoding and streaming delay, and low camera-side compute overhead).
We then elaborate on why prior solutions struggle to simultaneously meet the three requirements.



\tightsubsection{Video encoding for edge video analytics}

\begin{figure}
    \centering
    \includegraphics[width=0.95\columnwidth]{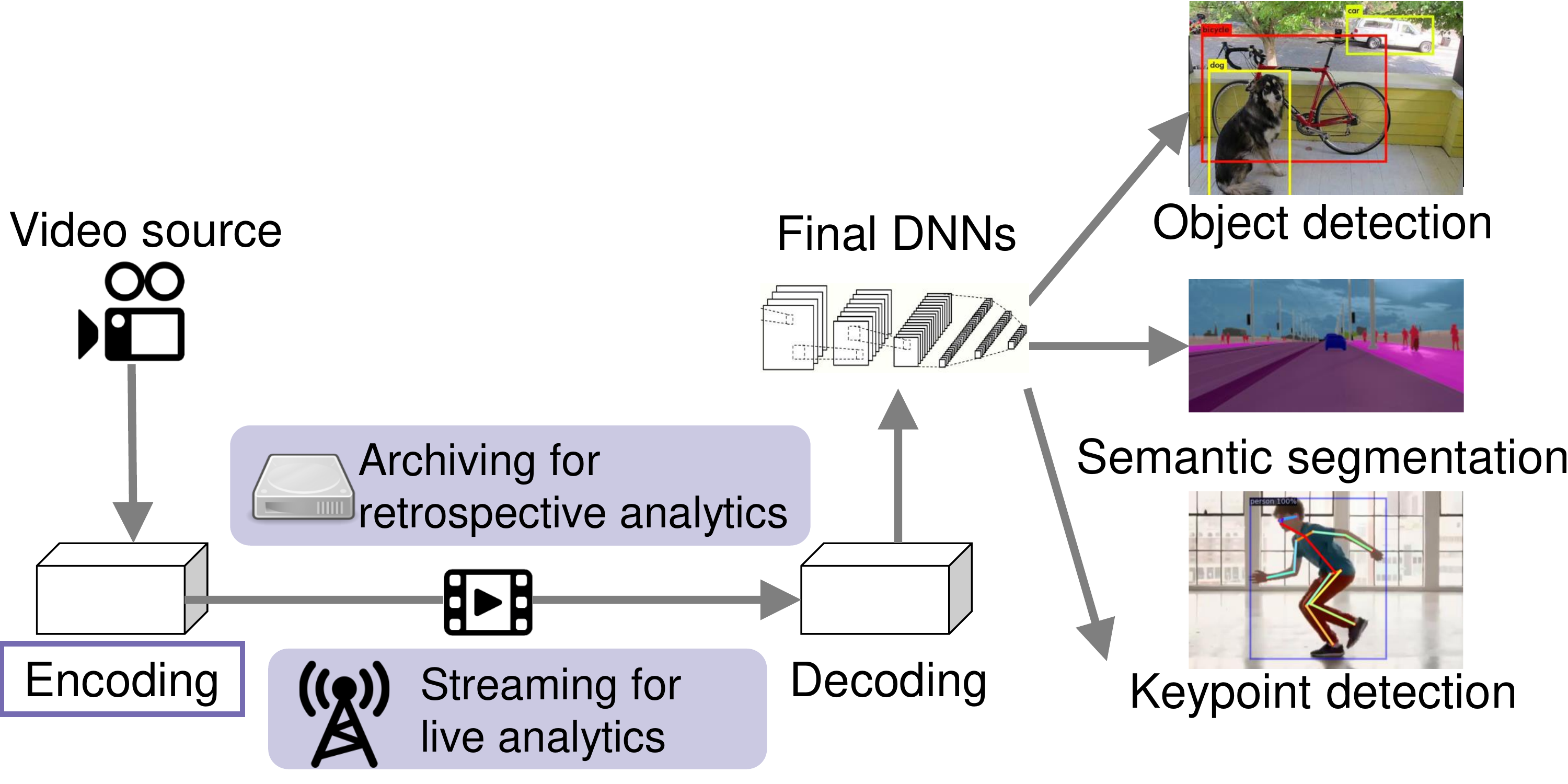}
    \tightcaption{\Edit Illustration of video encoding as part of the video analytics pipelines for three example tasks}
    \label{fig:background}
\end{figure}

\mypara{Distributed video analytics} 
As accurate analytics requires compute-intensive DNNs that cheap video sensors cannot afford, 
the video frames are often {\em compressed} by a {\em video encoder} and then sent to a remote server for accurate DNN-based analytics (Figure~\ref{fig:background}).
We refer to the server-side DNN as the {\em final DNN}.
In this work, we focus on three video analytics tasks: object detection (one labeled bounding box for each object), semantic segmentation (one label for each pixel), and keypoint detection (17 keypoints such as hand and elbow on a human body). 

There are two types of video analytics.
In live analytics, the video frames are continuously encoded and sent to a remote server which runs the final DNN to analyze the video in an online fashion~\cite{dds,eaar,vigil,glimpse,reducto,elf,awstream,noscope,videostorm}.
In retrospective analytics, the encoded video is first stored locally on the camera, and an operator can choose to fetch part of the video for DNN-based analytics~\cite{chameleon,blazeit,focus}.

\mypara{Performance requirements}
In both live and retrospective analytics, a key component is the video encoding algorithm.
An ideal video encoding algorithm for distributed video analytics should meet three goals:\footnote{
Different video analytics applications may have different objectives in terms of inference accuracy and delay (\eg augmented reality is more delay sensitive than traffic monitoring, and vehicle collision detection is more sensitive to accuracy than vehicle counting). 
Instead of evaluating performance of a video analytics pipeline in the context of a particular application, this work focuses on the relative improvement in accuracy, delay, and camera-side compute overhead, and we leave application-specific interpretation of the performance to future work.
} 
\begin{packeditemize}
\item {\em High accuracy:}
The encoded video must preserve enough information for the final DNN to return nearly identical inference results as if it runs on the original video frames.\footnote{To calculate the accuracy of an inference result on a compressed frame, we obtain the ``ground truth'' results by running the final DNN on the high-quality video frames (rather than using the human-annotated labels). 
Thus, any inaccuracy will be due to the video stack (\eg video compression, DNN distillation), rather than errors made by the final DNN itself.
This is consistent with recent work (\eg~\cite{awstream,vigil,videostorm,chameleon,mullapudi2019online,noscope}).}

\item {\em Low delay:} The delay of encoding the video (encoding delay) and streaming the video to the server (streaming delay) should be low.

\item {\em Low camera cost:} The encoding algorithm should be cheap enough to run at 30fps with only marginal extra compute overhead compared to encoding videos using popular codecs such as h.26x.
\end{packeditemize}


\tightsubsection{Limitations of previous work}
\label{subsec:challenge}

Here, we categorize previous work in four general approaches and explain why they cannot meet all performance requirements simultaneously.



\mypara{Local frame-filtering schemes}
One of the popular techniques is to let the camera run a simple logic to identify which frames are irrelevant to the vision task and thus can be discarded~\cite{reducto,filterforward,glimpse} or encoded in low quality~\cite{awstream}.
This approach works well when the video content is relatively stationary, where the incidents/objects of interest are rare and easy to detect; \eg in wildlife camera feeds, animals are rare and readily detectable since they are the only moving objects on a static background.
However, for frames that are not discarded, this approach encodes the entire frames with uniform quality, which can be suboptimal, since the objects of interest often occupy only a small fraction of each frame~\cite{dds,eaar}, leading to higher streaming delays than necessary.\footnote{Although CloudSeg~\cite{cloudseg} does not perform frame filtering at the camera side, it also shares the limitation since it compresses entire frames in same encoding quality.}

\mypara{Local heuristics to lower background quality}
Since objects of interest often account for a small fractions of each frame, some work (\eg~\cite{vigil,cinet}) uses local heuristics to filter out (or lower the quality of) the background pixels and 
sends the remaining object-related pixels in high quality to the server-side final DNN.
However, these local heuristics are constrained by the limited camera-side compute resources, giving rise to false negatives---object-related pixels are treated as background and thus filtered out or sent in low quality, causing the DNN to miss objects of interest.
For instance, to detect potential object-related regions, Vigil~\cite{vigil} relies on a low-accuracy non-convolutional Haar-cascade-classifier-based object detector, and CiNet~\cite{cinet} uses a very shallow convolutional network (with only 2 convolutional layers, 1 average pooling layer and 2 fully-connect layers) designed to handle only few objects per frame.
There are also deeper NNs such as MobileNet-SSD~\cite{ssd} that run on resource-constrained cameras, but they have to downsize frames to low resolutions (\eg 300$\times$300) for real-time inference, thus prone to missing small objects.


\mypara{Server-driven compression}
To overcome the camera-side resource constraints, another approach~\cite{dds,eaar,elf} leverages the abundant server-side compute resources to generate feedback on how videos should be encoded.
This approach generally compresses videos efficiently while achieving high accuracy, but it suffers from a high inference delay.
DDS~\cite{dds}, for instance, sends a low-quality video to the server-side DNN which returns to the camera which regions must be encoded in high quality, but under high network latency, getting such server-driven feedback can take at least two network round-trip times before the camera can actually encode the video for final DNN inference, causing high inference delay on each frame.

\begin{figure*}[!t]
\centering
    \includegraphics[width=1.7\columnwidth]{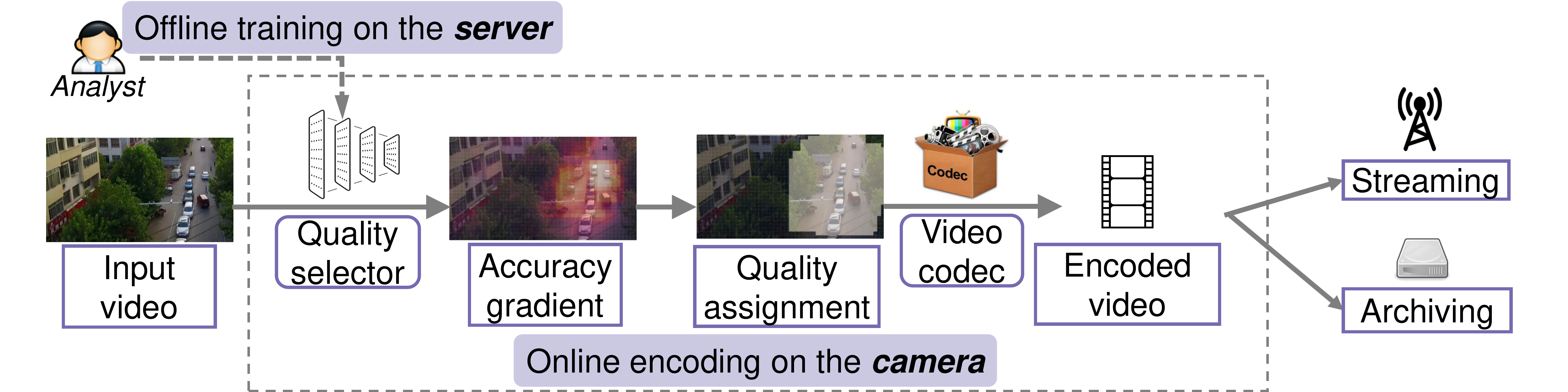}
    \tightcaption{\Edit Overview of \name.}
    \label{fig:workflow}
\end{figure*}

\mypara{Local DNN compression}
Instead of encoding videos on the camera, some proposals also extract the final DNN's feature maps on the camera and compress the feature maps which might contain less information than the original raw frames~\cite{duan2020video,xia2020emerging,cracking,neurosurgeon,matsubara2019distilled}. 
While this approach has shown promise with classification or action recognition DNNs~\cite{neurosurgeon,cracking,duan2020video}, these tasks do not require the spatial locations of objects, allowing aggressive aggregation of feature maps over an entire frame. 
In contrast, the vision tasks that we focus on (\eg object detection, semantic segmentation) are sensitive to object locations, making the intermediate feature maps much larger and much more difficult to be compressed efficiently. 
For example, many state-of-the-art object detectors (\eg~\cite{detectron2}) use expensive feature extractors such as ResNet101~\cite{resnet}, and if we feed a 720p ($1280\times720$) frame through even parts (\eg 90) of its convolution layers, the feature map still contains $2\times 10^7$
floating-point numbers per frame, 20x more than the number of pixels in the original frame.

There are also proposals to train DNN autoencoders that compress video to a smaller size than the popular video codecs do, but these DNNs are much more compute-intensive than the video codecs and even the final DNN.
For example, NLAM~\cite{NLAM} requires performing expensive 3D convolutions on videos for more than 30 times, while an object detector backbone (\eg ResNet34~\cite{resnet}) only performs 2D convolution for 34 times.

\tightsection{Overview of \name}
\label{sec:overview}

In this section, we present \name, a new video encoding algorithm that uses a {\em cheap} camera-side model to decide which regions should be encoded in higher quality. 
Here, we introduce \name's workflow and its challenges, and then present the key idea that addresses these challenges.

\tightsubsection{Workflow and challenges of \name}
\label{subsec:workflow-challenge}

Figure~\ref{fig:workflow} depicts the workflow of \name.
    When a video frame arrives, \name first feeds it through a cheap quality selector model, called \AccuracyGradientModel, to obtain a {\em macroblock-level quality selection}---which macroblocks (16x16 blocks, which many modern video codecs~\cite{h264} use as the basic encoding unit) should be encoded in high quality and which should be in low quality.
    The camera then encodes the video frames according to the quality selection and sends the encoded video to the server for the final DNN inference.
    The quality selector (\AccuracyGradientModel) is trained for each final DNN offline, such that when the video frames are encoded in its selected quality, the DNN can return accurate inference results. As we will see in \S\ref{sec:training}, the quality selector can also be re-used among DNNs of similar tasks with only marginal performance penalty.

\mypara{Challenges} 
The key component of \name is the quality selector (\AccuracyGradientModel) that selects the encoding quality per macroblock.
It has three challenges:
{\em (i)} How to optimally assign encoding quality at the fine spatial granularity of macroblocks to achieve better accuracy-delay tradeoffs than baselines?
{\em (ii)} How to minimize the per-frame compute overhead of \AccuracyGradientModel to allow real-time video encoding on the camera side?
And {\em (iii)} how to quickly train \AccuracyGradientModel for each server-side DNN?

\tightsubsection{Key idea: Accuracy Gradient}
\label{subsec:accuracy-gradient}

The key idea to address these challenges of \AccuracyGradientModel is to obtain the {\em Accuracy Gradient}  (hereinafter \AccuracyGradient) of each macroblock, which measures how much the encoding quality at the macroblock can affect the DNN inference accuracy.
Mathematically, the \AccuracyGradient of macroblock \Block in a given frame is defined as:

\vspace{-0.6cm}
{
\footnotesize
\begin{align}
\AccuracyGradient_{\Block}
= \sum_{\Pixel\in \Block}
\bigg\Vert
\frac{\partial \AccuracyFunc\big(\DNN(\Image);\DNN(\HighImage))}{\partial \Image_{\Pixel}}\bigg|_{\Image=\LowImage}\bigg\Vert_1 \cdot \Vert \HighImage_\Pixel - \LowImage_\Pixel\Vert_1, \label{eq:acc-gradients}
\end{align}
}
\vspace{-0.8cm}

\noindent where \Pixel, \LowImage, and \HighImage denote a pixel within \Block, the low-quality encoded frame, and the high-quality encoded frame, respectively.
\DNN denotes the server-side DNN inference function, \AccuracyFunc$(\DNN(\Image);\DNN(\HighImage))$ is the accuracy of inference result on $\Image$ (\ie its similarity with the inference result on the high-quality frame $\DNN(\HighImage)$),
and $\Vert\cdot\Vert_1$ is the L1-norm.

Intuitively, the encoding quality of the macroblocks with higher \AccuracyGradient values have a greater influence on the DNN accuracy, so these macroblocks should be encoded in high quality.
We will present the \AccuracyGradient-based quality assignment logic in \S\ref{sec:encoding}.
Appendix~\ref{appendix:accuracy-gradient} gives the mathematical reasoning behind Equation~\ref{eq:acc-gradients}.

\AccuracyGradient enables a series of system optimizations that help address the challenges in \S\ref{subsec:workflow-challenge}.



\begin{figure}
    \centering
    \includegraphics[width=0.95\columnwidth]{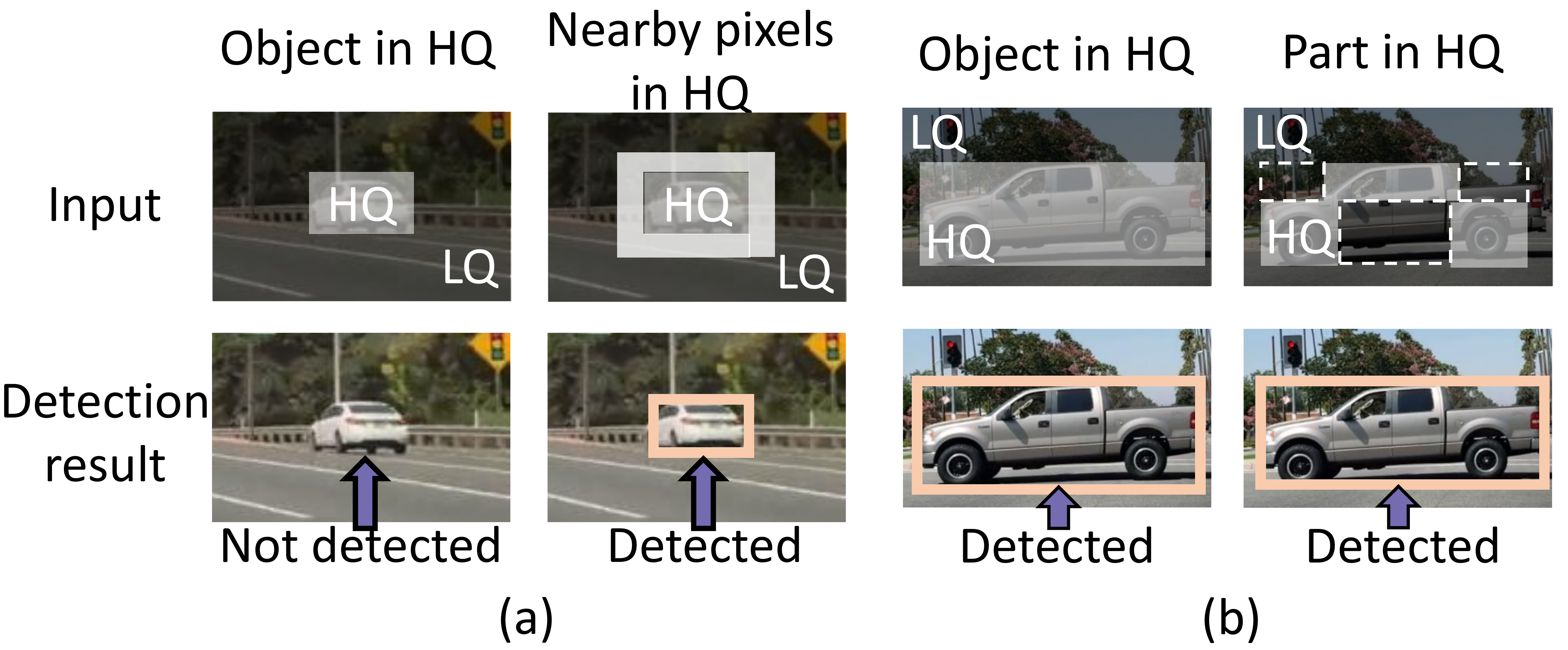}
    \vspace{-0.2cm}
    \tightcaption{
    Examples of inefficiencies of object-based encoding (high quality in the object bounding box):
    (a) Objects not detected unless nearby pixels are in high quality; and (b) Objects still detected even if just parts are in high quality.
    }
    \vspace{-0.1cm}
    \label{fig:object-hqblock}
\end{figure}

\red{\mypara{Advantage over prior region-based compression}}
We begin with the benefit of \AccuracyGradient-based quality selection over the traditional region-based quality selection~\cite{dds,eaar,elf}, which identifies regions with greater impact on DNN inference (\eg via region proposals~\cite{faster-rcnn}) and then use high quality to encode some region proposals in their {\em entirety} and low quality to encode the {\em whole} background. 
These coarse-grained encoding schemes can be inefficient. 
On one hand, some surrounding pixels of the object bounding boxes can still be crucial for the final DNN to accurately detect/classify objects and demarcate their boundaries from the background.
For instance, to detect the car in Figure~\ref{fig:object-hqblock}(a), one must encode not only its bounding box in high quality but also some neighboring macroblocks too.
On the other hand, some pixels inside an object's bounding box (\eg the smooth surface of a car in Figure~\ref{fig:object-hqblock}(b)) have similar RGB values regardless of the encoding quality, so it is safe to compress them in low quality without hurting inference accuracy.

In contrast, \AccuracyGradient by definition can capture such fine distinctions among macroblocks.
For instance, the macroblocks surrounding the car's bounding box (Figure~\ref{fig:object-hqblock}(a)) will have high \AccuracyGradient values in Equation~\ref{eq:acc-gradients}, because $\AccuracyFunc\big(\DNN(\LowImage);\DNN(\HighImage))$ will have a high derivative with respect to the pixels in these macroblocks. 
Similarly, the smooth surface of the car in Figure~\ref{fig:object-hqblock}(b) is likely to have low \AccuracyGradient (despite being part of the car object), because $\Vert\HighImage_\Pixel-\LowImage_\Pixel\Vert_1$ will be small on the pixels \Pixel in these macroblocks.\footnote{\AccuracyGradient may look similar to the saliency maps in computer vision, but there is a key distinction. 
While saliency captures which pixel {\em values} have more influence on the DNN output, \AccuracyGradient captures how much changing a macroblock's {\em encoding quality} changes the
DNN inference accuracy.}

\mypara{\AccuracyGradientModel might not be compute-intensive}
\AccuracyGradientModel can be seen as a segmentation problem, but unlike normal segmentation models that are compute-intensive, a cheaper model might suffice for \AccuracyGradientModel for three reasons.

First, unlike traditional image segmentation that gives one label per pixel, \AccuracyGradientModel returns one label per {\em 16x16 macroblock} (\ie all pixels in a macroblock share the same DNN output). 
Therefore, unlike traditional convolutional operations which scan the image pixel by pixel, \AccuracyGradientModel only needs to scan the image {\em macroblock by macroblock} (saving upto $16^2=256$x on convolutional operations).

Second, \AccuracyGradientModel only needs to do a {\em binary} classification on each macroblock (either high quality or low quality), rather than multi-class segmentation, which further reduces the complexity of \AccuracyGradientModel. 

Third, while accurate segmentation must  minimize both false positives (\eg pixels misclassified as objects) and false negatives (\eg pixels misclassified as background), \name has more tolerance towards false positives (\eg macroblocks mislabeled with high \AccuracyGradient). Encoding a few more macroblocks in high quality has marginal impact on the delay-accuracy tradeoff, because the intra-frame encoding commonly used in video codecs makes the compressed video size grow only sublinearly with more high-quality regions. 
Appendix~\ref{appendix:false-positive-tolerance} provides the empirical evidence.

\mypara{Fast training of \AccuracyGradientModel}
Training \AccuracyGradientModel naively can be prohibitively expensive, because it requires running numerous forward/backward propagations on the final DNN to calculate losses and obtain gradients.
Fortunately, \AccuracyGradient can be directly derived from the final DNN (Equation~\ref{eq:acc-gradients}) using only two forward propagations and one backward propagation. 
As we will see in \S\ref{sec:training}, this allows us to separate the compute-intensive final DNN from \AccuracyGradientModel training. 
This can save 10$\times$ overhead, as 
\name trains \AccuracyGradient for 15 epochs (see \S\ref{sec:training} for details), each requiring only three propagations through the final DNN.

\tightsection{Online encoding}
\label{sec:encoding}


We now describe \name's online encoding process, including the architecture of \AccuracyGradientModel
and how \name assigns encoding quality to each macroblock.

\mypara{Architecture of \AccuracyGradientModel}
We leverage the pretrained MobileNet-SSD feature extractor~\cite{ssd}, a widely-used feature extractor for cheap edge devices, as the feature extractor of \AccuracyGradientModel.
We resize these features so that one macroblock corresponds to one feature vector, and append three convolution layers to classify which macroblock should be in high quality.



\mypara{Compute cost of \AccuracyGradientModel} 
Our model is much more compact than other commonly-used feature extractors.
To put it into perspective, our \AccuracyGradientModel uses 12 GFLOPs, about $3\times$ less than a typical cheap convolutional model such as ResNet18 which uses 33 GFLOPs.
In addition, since the architecture consists only of convolutional layers (except for batch normalization and activation), its computational overhead is proportional to the size of the input frame (\eg $4\times$ faster when the frame size halved in both dimensions).

\mypara{\AccuracyGradient-based quality assignment}
Given macroblock-level \AccuracyGradient of a frame, \name then uses a threshold $\AgThresh$ to determine which macroblocks should be in high quality---all blocks $\Block$ with $\AccuracyGradient_{\Block}\geq\AgThresh$ will be encoded in high quality.
After a set of blocks are selected, \name then 
expands these selected blocks to each direction by $\PadThresh$ (by default 5) blocks (if they are not already selected).
Intuitively, 
a lower $\AgThresh$ increases the accuracy at the expense of encoding more macroblocks with high \AccuracyGradient.

\begin{figure}[t]
    \centering
    \includegraphics[width=0.75\columnwidth]{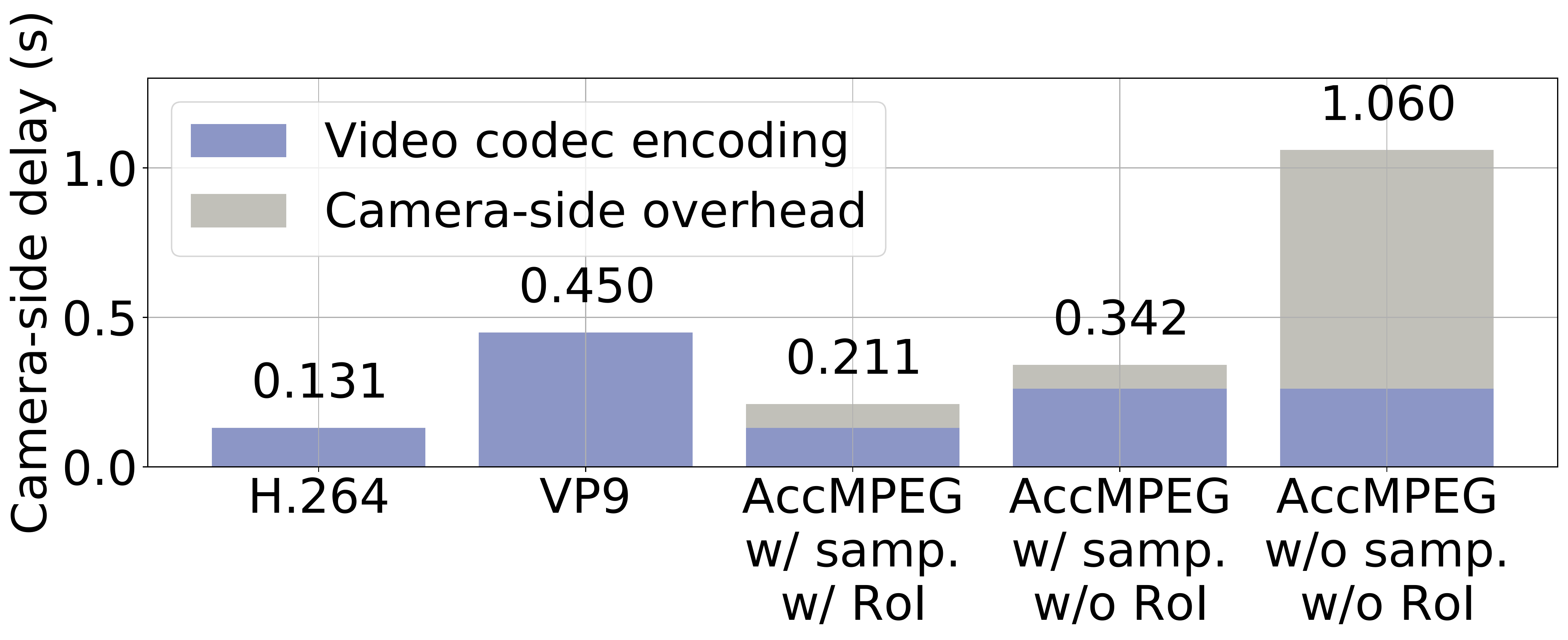}
    \tightcaption{Breakdowns of camera-side delay: the delay of running \AccuracyGradientModel is marginal (more so after the frame sampling optimization), compared to encoding delay of H.264 and VP9.
    }
    \vspace{0.1cm}
    \label{fig:eval-cost-reduction}
\end{figure}

\mypara{Frame sampling for cheap inference} 
We further reduce \AccuracyGradientModel's compute overhead by running it once every $k$ frames and using its output to encode the next $k$ frames (by default, $k=10$).
Empirically, it significantly reduces the camera-side overhead (Figure~\ref{fig:eval-cost-reduction}) without much impact on accuracy. 
The intuition is that although the \AccuracyGradient of a macroblock fluctuates over time, its value will not shift dramatically across consecutive frames to change its quality selection.

\begin{figure}[t]
    \centering

    \includegraphics[width=0.6\columnwidth]{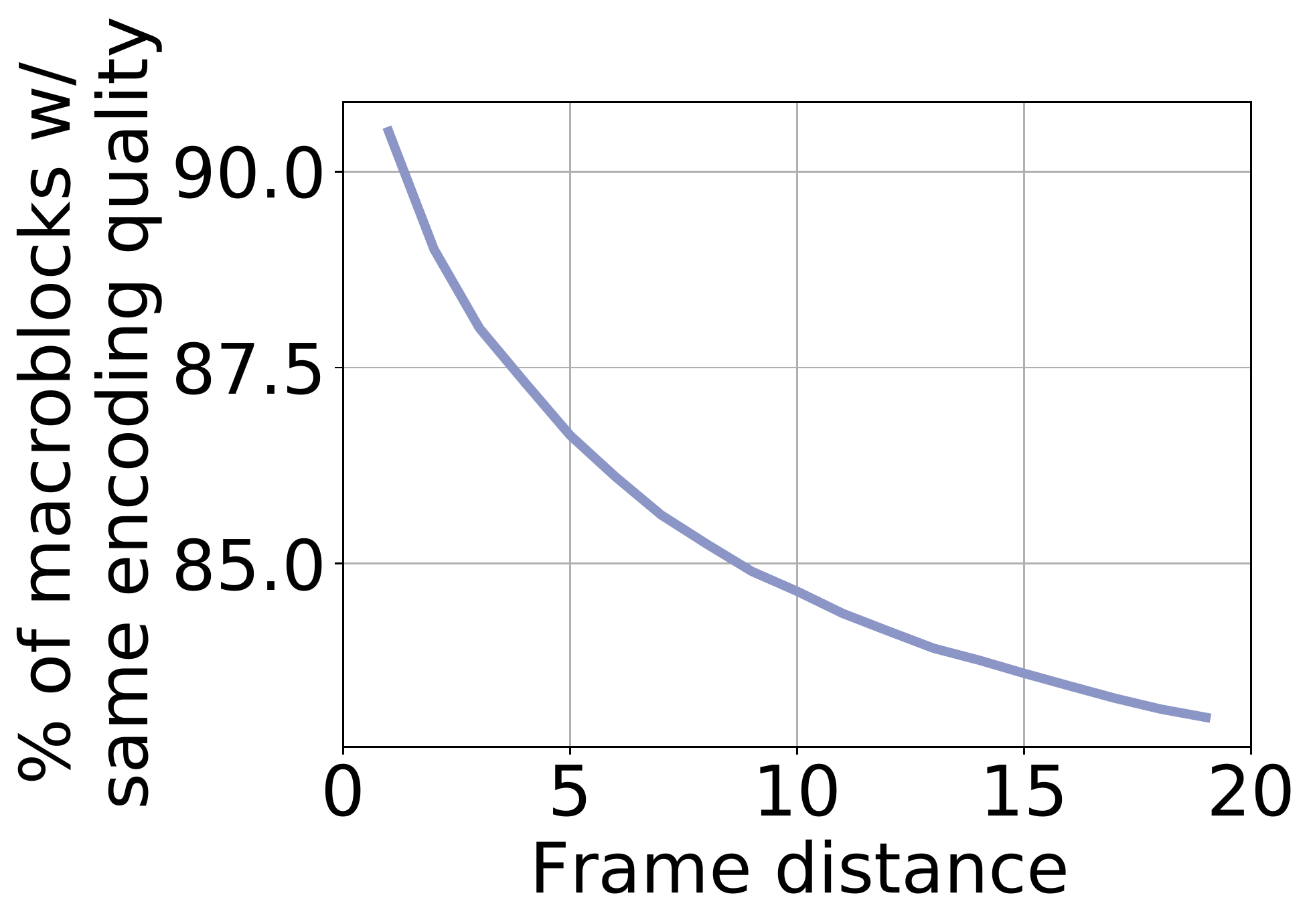}
    \tightcaption{\Edit The encoding quality of over 84\% of the macroblocks remains to be the same 
    when the frame distance is less than 10.}
    \label{fig:assignment-persistence}
\end{figure}

{
\Edit
To empirically verify this intuition, we use frame distance (the absolute delta between the frame numbers of two frames) to measure the temporal distance between two frames and examine how the encoding quality assignment (we measure this change by the percentage of macroblocks that the encoding quality remains the same) varies with greater frame distance.
Figure~\ref{fig:assignment-persistence} shows that on the quality assignment generated by running \AccuracyGradientModel on images of the dashcam dataset (see \S\ref{sec:eval} for the detail of the dataset),
the encoding quality assignment of at least 84\% of macroblocks remains unchanged within 10 consecutive frames.

}

\tightsection{Offline training}
\label{sec:training}

\begin{figure}
    \centering

    \subfloat[\centering{Conventional training pipeline}]
    {
        \includegraphics[width=1\columnwidth]{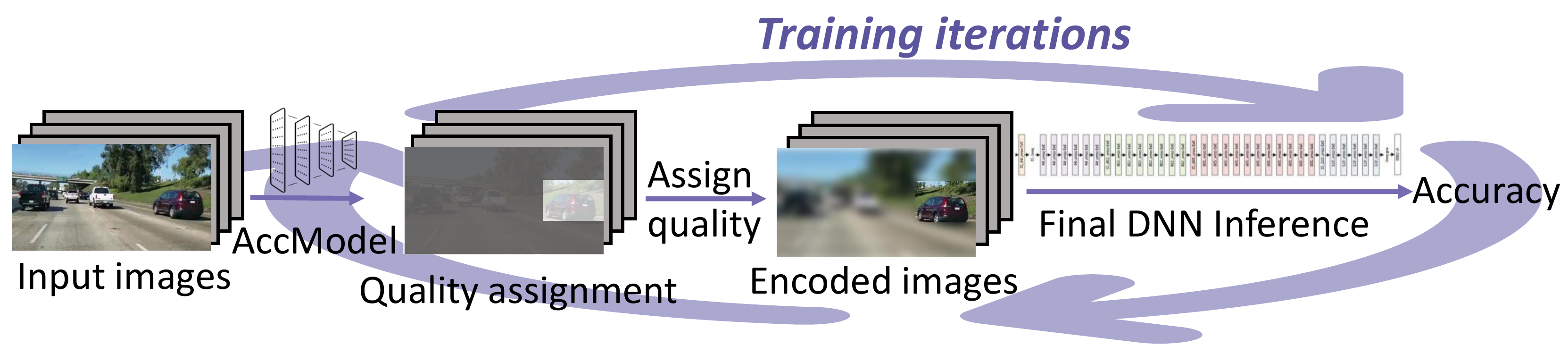}
        \label{fig:training-conventional}
    }\vspace{-0.35cm}
    \subfloat[\centering{Decoupling the final DNN from training using \AccuracyGradient}]
    {
        \includegraphics[width=0.7\columnwidth]{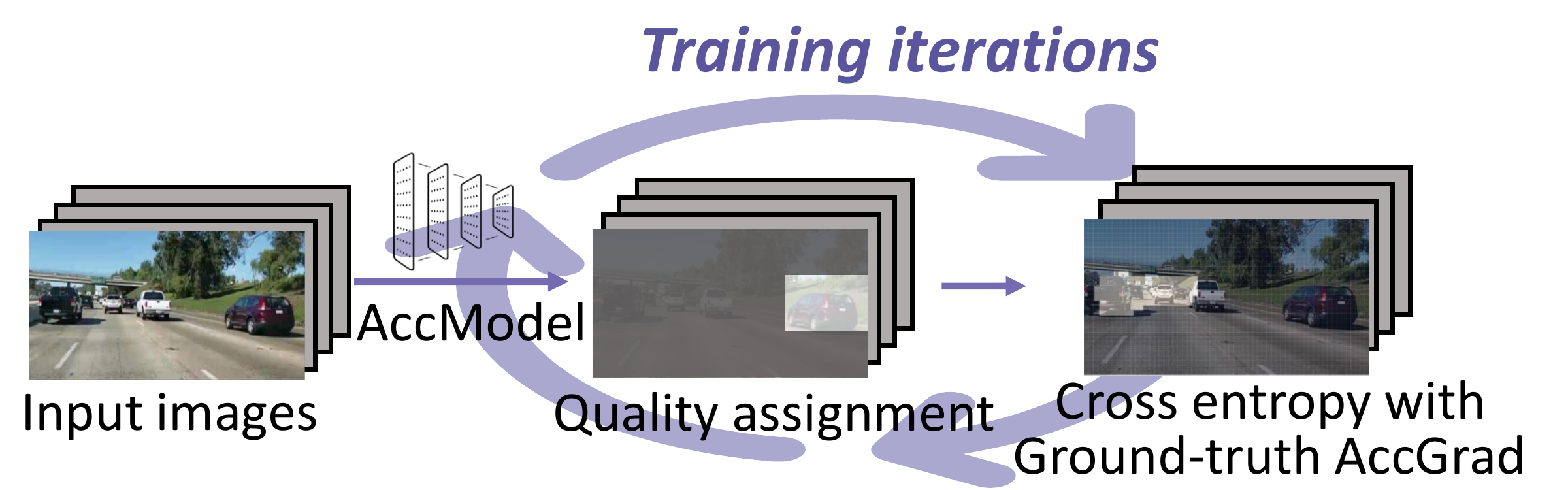}
        \label{fig:training-new}
    }
    \tightcaption{Contrasting the conventional  approach to \AccuracyGradientModel training with \name.
    We separate the final DNN from training by first generating the ground-truth \AccuracyGradient from the final DNN, and then training \AccuracyGradientModel to minimize the loss with the ground-truth \AccuracyGradient, which significantly speeds up the training.
    }
    \label{fig:training}
\end{figure}

\AccuracyGradientModel customizes the video encoding for near-optimal accuracy-delay tradeoffs of a given final DNN model. 
So a natural question is how to quickly create the \AccuracyGradientModel for a new DNN.
Here, we describe how we speed up the training of \AccuracyGradientModel by separating the final DNN from the training process, and then explain why reusing \AccuracyGradientModel may also lead to decent performance. 






\mypara{Conventional training process}
Before describing how \name trains \AccuracyGradientModel, we first explain the straightforward approach to training \AccuracyGradientModel (depicted in Figure~\ref{fig:training-conventional}) which sets up the entire pipeline of encoding and inference and minimizes the end-to-end loss (we will define the training loss soon).
For each input image, it first feeds the high-quality version $\HighImage$ through the \AccuracyGradientModel to get the \AccuracyGradient matrix $\Mask=\AccuracyGradientModel(\HighImage)$, creates the encoded image $\Image=\Mask\times\HighImage+(\AllOnes-\Mask)\times\LowImage$ by linearly combining the low-quality version $\LowImage$ and $\HighImage$ using $\Mask$\footnote{We make the elements in $\Mask$ to be between 0 and 1 by applying a softmax filter on the output of \AccuracyGradientModel.}, 
feeds it through the final DNN to get result $\DNN(\Image)$, and finally calculates the accuracy (loss) of $\DNN(\Image)$ with $\DNN(\HighImage)$ as the ground truth using $\AccFunc$. 
This training process is actually widely used in computer vision (\eg~\cite{GAN,perceptual-loss,SRGAN}) and video analytic systems~\cite{cloudseg}.
However, it is prohibitively expensive: each forward or backward propagation of the pipeline must run the expensive $\DNN$, in addition to \AccuracyGradientModel and $\AccFunc$ each once.
Thus, the compute overhead of \AccuracyGradientModel training is dominated by running the forward/backward propagations on $\DNN$ (caching \AccuracyGradientModel results does not help, since the \AccuracyGradientModel output changes after each update).

\mypara{Separating final DNNs from training via \AccuracyGradients}
In contrast, \name calculates the \AccuracyGradients on each training image first (using Equation~\ref{eq:acc-gradients}), which requires only one forward and backward propagation of $\DNN$ on the high-quality version of the image. 
Once the \AccuracyGradients of each training image is generated, the training can be reformulated as $\min_{\AccuracyGradientModel} CrossEntropy\big(\AccuracyGradientModel(\HighImage), \Mask^*\big)$,
where $\Mask^*$ is the ``ground truth'' \AccuracyGradient matrix of size $\GridWidth\cdot\GridHeight$ generated by Equation~\ref{eq:acc-gradients} (not to be confused with the ground-truth inference output of a final DNN) and the cross-entropy loss (with 4x weight on those blocks that should be in high quality) commonly used in deep learning to measure the discrepancies between two vectors.
Training \AccuracyGradientModel thus requires only one forward and one backward propagation on \AccuracyGradientModel. 
Thus, by generating the ground truth first and then training \AccuracyGradientModel, we can train the \AccuracyGradientModel within 8 minutes using 8 GPUs (\S\ref{subsec:eval:customization}).

\mypara{Using pre-trained models}
Instead of training  \AccuracyGradientModel from scratch, we initialize \AccuracyGradientModel with a pretrained MobileNet-SSD backbone and then fine-tune the model.
It has the similar benefit of model fine-tuning widely used in industry: the training can converge with fewer training epochs on fewer training images~\cite{gao2021recent}. 
Specifically, we train \AccuracyGradientModel on a 10$\times$ randomly downsampled training set of the final DNN model (\eg COCO dataset) for 15 training epochs and pick the model with lowest loss on cross validation set as our final \AccuracyGradientModel. 
The total training time of \AccuracyGradient is about 8 minutes (\S\ref{subsec:eval:customization}).

\begin{figure}[t]
    \centering
    \subfloat[ ][\vbox{\centering{FasterRCNN$\rightarrow$EfficientDet\\(Highway)}}]
    {
        \includegraphics[width=0.45\columnwidth]{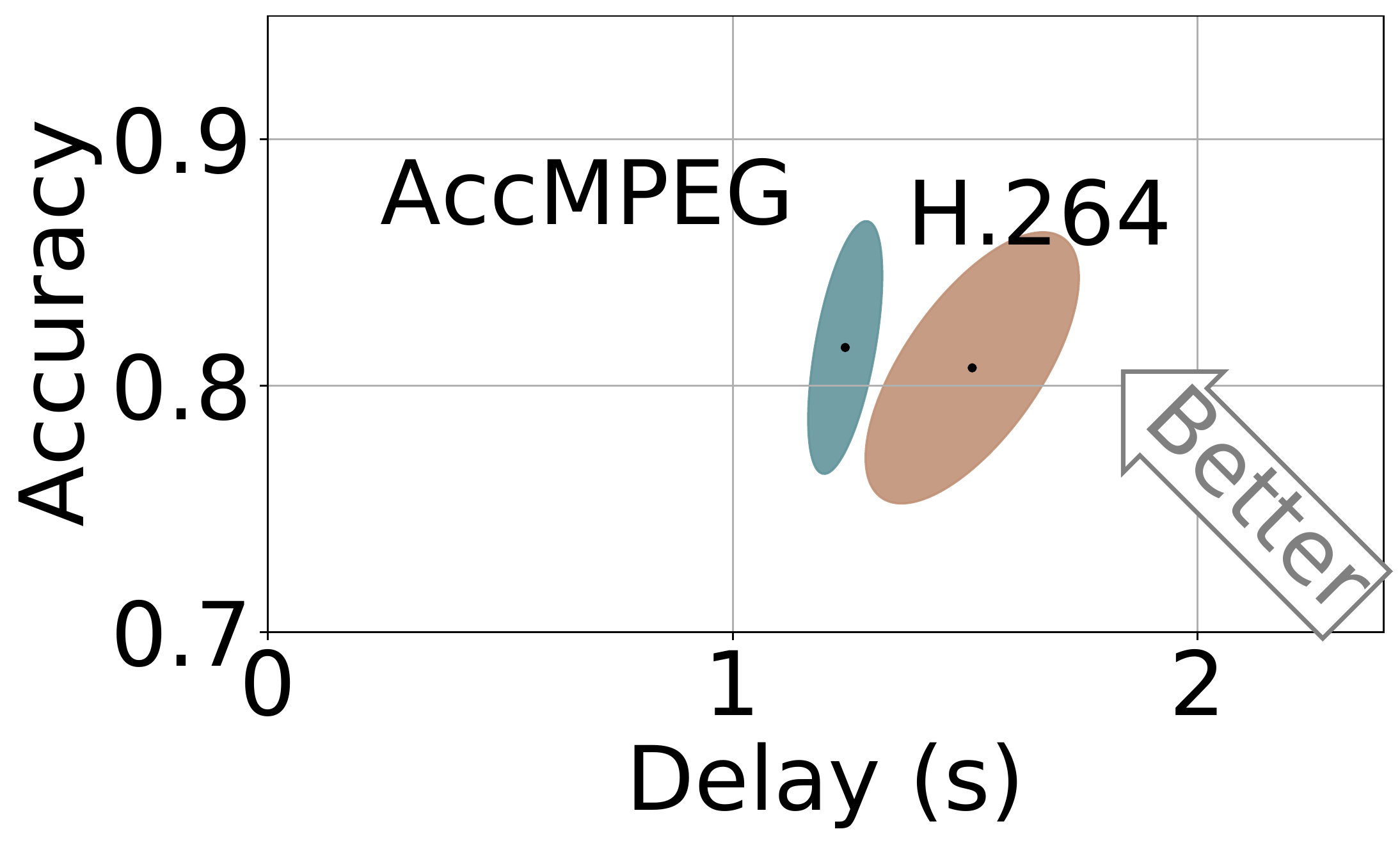}
        \label{subfig:1}
    }
    \subfloat[ ][\vbox{\centering{FasterRCNN$\rightarrow$YoLo\\(Highway)}}]
    {
        \includegraphics[width=0.45\columnwidth]{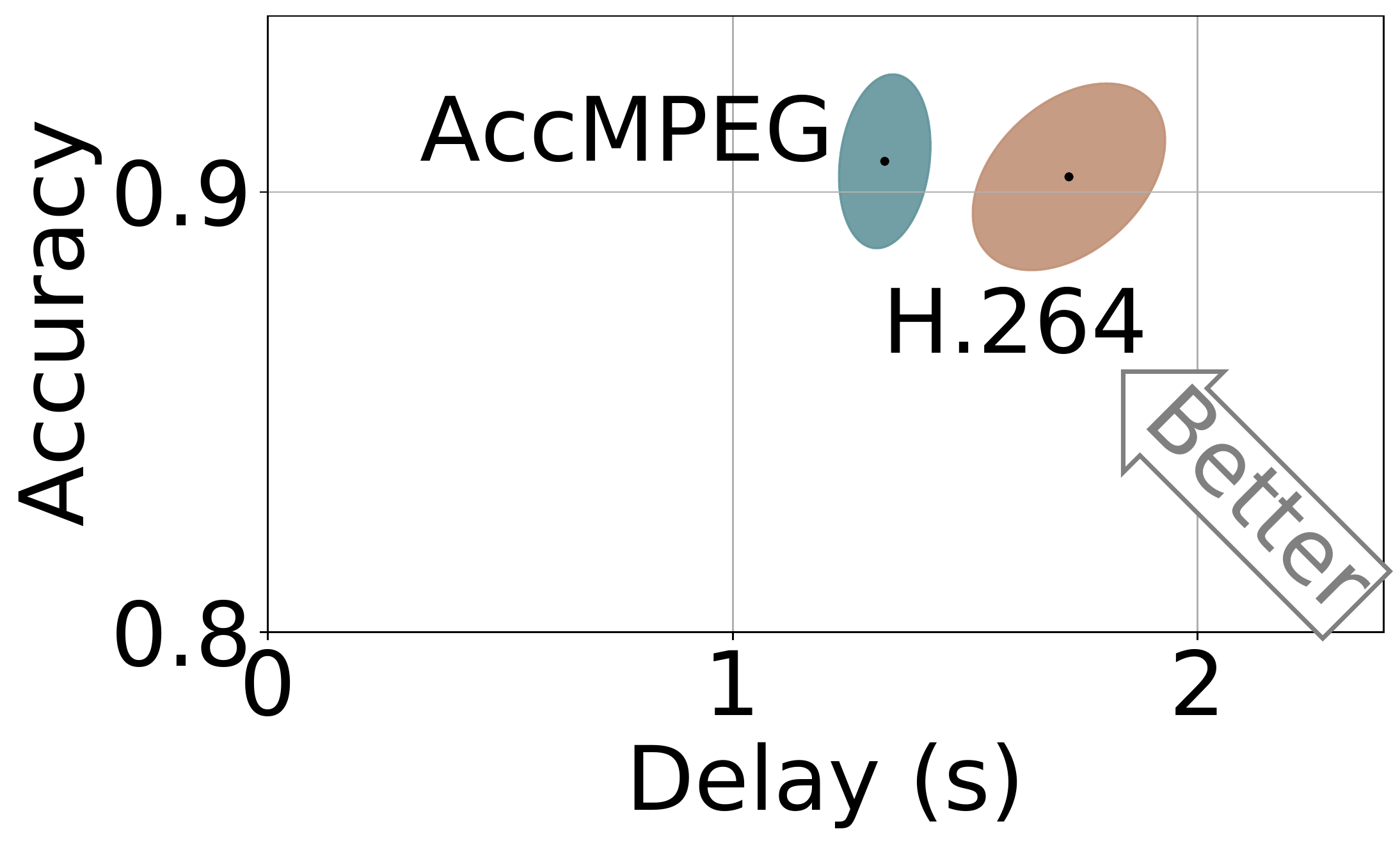}
        \label{subfig:1}
    }
    
    \vspace{-0.4cm}
    \subfloat[ ][\vbox{\centering{FasterRCNN$\rightarrow$EfficientDet\\ (Driving)}}]
    {
        \includegraphics[width=0.45\columnwidth]{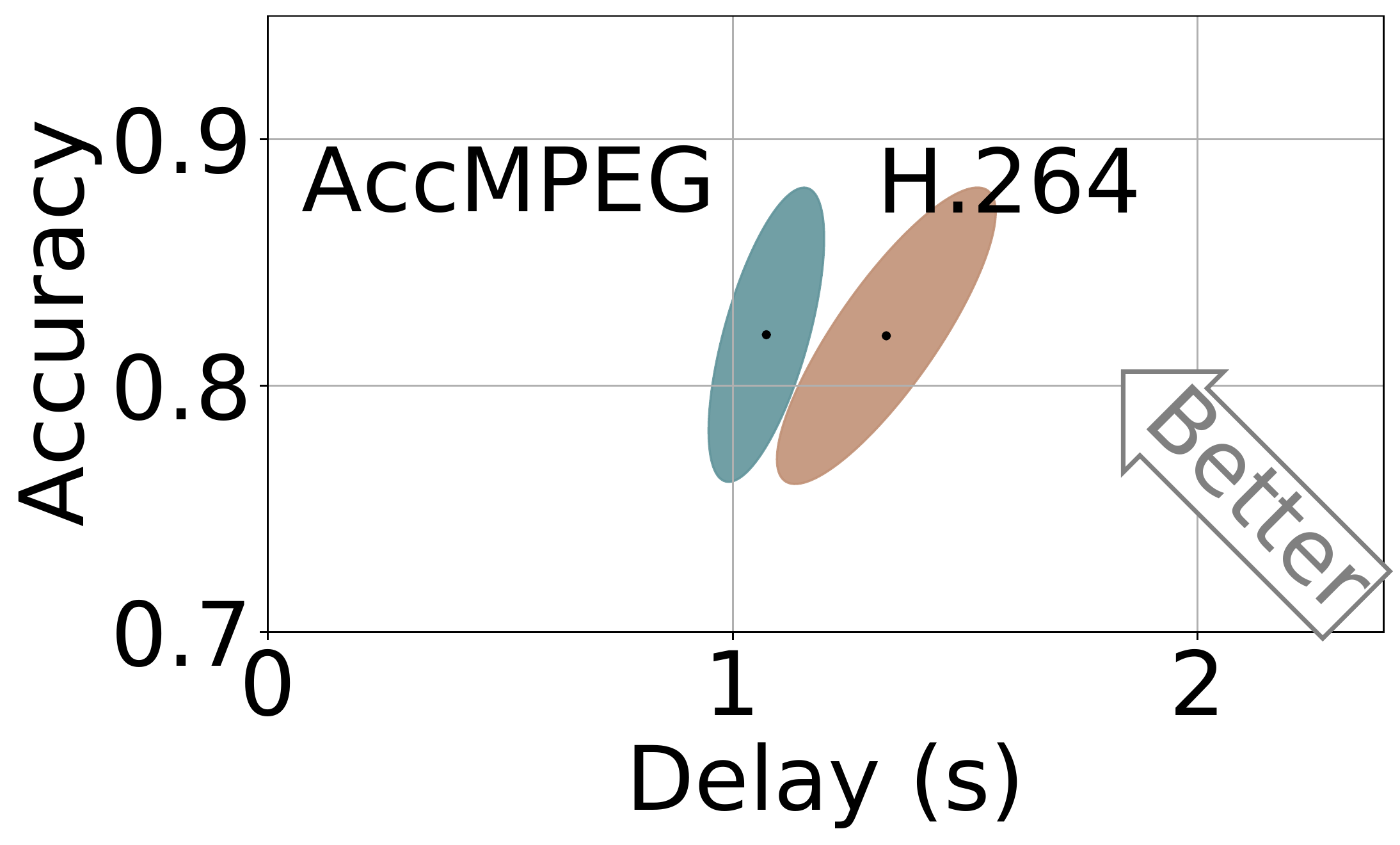}
        \label{subfig:1}
    }
    \subfloat[ ][\vbox{\centering{FasterRCNN$\rightarrow$YoLo\\ (Driving)}}]
    {
        \includegraphics[width=0.45\columnwidth]{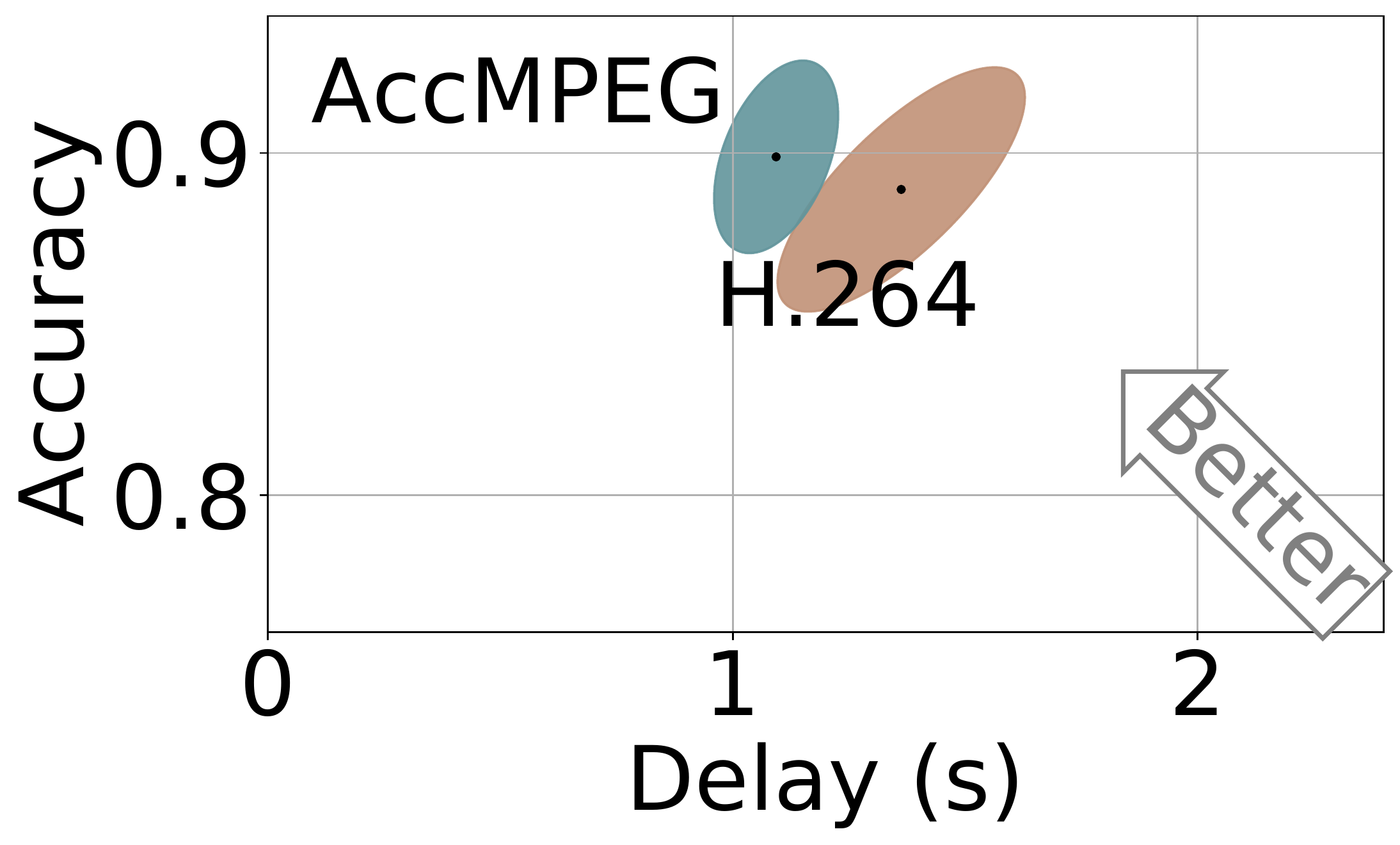}
        \label{subfig:1}
    }
    
    
    \tightcaption{Even if we reuse the \AccuracyGradientModel trained for a different final DNN, \name still offers decent performance gain over the best H.264 encoding scheme. 
    DNN $A\rightarrow$ DNN $B$ means the \AccuracyGradientModel trained for $A$ is reused to encode videos for $B$.
    }
    \label{fig:pretrain}
\end{figure}

\mypara{Reusing \AccuracyGradientModel}
Ideally, any new server-side final DNN requires a (slightly) different \AccuracyGradientModel.
However, when the new final DNN is trained on the same dataset (same images and same labels) as another final DNN (whose \AccuracyGradientModel is already trained), it is possible to reuse the \AccuracyGradientModel. 
This is because the macroblocks with high \AccuracyGradients are typically those related to small, partially occluded, or darkly lit objects in the dataset.
Thus, training the new \AccuracyGradientModel based on the \AccuracyGradients of the old final DNN on the same dataset would likely yield a similar \AccuracyGradientModel.
Since DNN models are sometimes trained on popular datasets (such as the COCO dataset~\cite{coco}), \AccuracyGradientModel can sometimes be re-used among different final DNNs (we will empirically evaluate it in Figure~\ref{fig:pretrain}).

\tightsection{Evaluation}
\label{sec:eval}

Finally, our evaluation of \name shows that:
\begin{packeditemize}
\item \name achieves better accuracy-delay tradeoffs: 10-43\% lower delay while maintaining comparable accuracy as the baselines. 
The improvement remains similar on three vision tasks and five final DNN models with a variety of architectures and backbones (\S\ref{subsec:eval:e2e}).
\item \name has the lowest camera-side overhead compared to all the baselines that deploy customize logic at the camera side and achieve comparable accuracy, and the extra compute overhead due to \AccuracyGradientModel is less than the popular video codecs
(\S\ref{subsec:delay}).
\item Given a final DNN, an \AccuracyGradientModel can be created within 8 minutes using 8 GPUs. 
Even if a final DNN changes without updating the  \AccuracyGradientModel, \name still achieves better accuracy-delay tradeoffs if the new vision model is trained on the same dataset as the previous one (\S\ref{subsec:eval:customization}). 
\end{packeditemize}

\tightsubsection{Setup}
\label{subsec:eval:setup}

\begin{table}
{\small
\begin{tabular}{c>{\centering\arraybackslash}m{3.1cm}cc} 
\hline
Name & Vision task & \# Videos & \# Frames \\ [0.5ex]
\hline
\textsf{Driving} & object detection & 5 & 9000 \\ 
\hline
\textsf{Dashcam} & object detection & 7 & 12600 \\
\hline
\textsf{Surf} &  \begin{tabular}[x]{@{}c@{}}semantic segmentation\\keypoint detection\end{tabular} & 6 & 6598 \\
\hline
\end{tabular}
\tightcaption{Summary of our datasets.
}
\label{tab:dataset}
}
\end{table}

{
\Edit
\mypara{Source code}
Our source code is publicly available (see appendix \ref{appendix:AE} for the details).
}

\mypara{Dataset} 
Table \ref{tab:dataset} summarizes the 3 video datasets we used to evaluate \name: 5 driving videos and 7 dashcam videos for object detection, and 6 surfing videos for keypoint detection and semantic segmentation. All videos are obtained by searching on YouTube. We search keywords (such as ``highway dashcam hd'') in incognito mode to avoid customization bias.
All videos and collection details are available in this anonymous link~\cite{dataset-link}.

\mypara{Device setting}
We create a 30fps video source where different methods can read raw (1280$\times$720) frames one by one. 
To achieve real-time video streaming, we let the camera stream out the video in the form of short video chunks (this aligns with previous work~\cite{dds,awstream}), each consisting of 10 frames.
To fairly compare the encoding delay of different methods, we benchmark the encoding delay on one Intel Xeon Silver 4100 CPU and run the encoding of \name and baselines everytime the camera reads 10 frames for its current video chunk (we also benchmark the performance of \name on baselines on  Jetson Nano, a cheap GPU device (with one 128-core Maxwell GPU, one Quad-core ARM A57 CPU and 4GB memory~\cite{chameleon-hardware}))\footnote{\label{footnote:cheap-jetson-nano}A Jetson nano developer board is only 60\$~\cite{jetson-nano-price}.} provided in the Chameleon testbed~\cite{chameleon-testbed}).
We use openVINO to accelerate\footnote{This acceleration will not reduce floating point precision, and thus will not alter the inference result of \name.} all camera-side DNNs on CPUs.
We also make minor modifications to the H.264 codec to 
enable macroblock-level region-of-interest encoding.

\mypara{Server}
We train \AccuracyGradientModel offline on the server with 8 GeForce RTX 2080 SUPER GPU.
In the online encoding phase, we run the decoding on Intel Xeon Silver 4100 CPU and run the inference on  GeForce RTX 2080 SUPER GPU. 


\mypara{Video analytics tasks and DNNs}
We test \name on three tasks: object detection, semantic segmentation and keypoint detection. Here we list the DNNs we use for these tasks (We use {\em italic} to show the DNN that we use to deliver \AccuracyGradientModel for that vision task. All DNNs are pretrained from COCO dataset~\cite{coco}).
    We pick three {\em object detection} models that represent three types of different architectures: \textit{FasterRCNN}~\cite{faster-rcnn} (a two-stage detector with features from different resolutions~\cite{fpn}), { YoLov5}~\cite{yolo} (a single-stage detector), and { EfficientDet}~\cite{tan2020efficientdet} (a detector with machine-optimized architecture~\cite{nas}).
We pick \textit{FCN-ResNet50}~\cite{fcn-resnet101} for {\em semantic segmentation}.
We also pick two {\em keypoint detection} models: \textit{Keypoint-ResNext101}~\cite{maskrcnn,resnext} and Keypoint-ResNet50~\cite{maskrcnn},

\mypara{Setting of \name}
For the encoding quality, we use (30, 40) as the QP value for high quality and low quality for object detection and (30, 51) for keypoint detection.
By default, we use $\alpha=0.2$ as the \AccuracyGradient threshold.

\mypara{Baselines}
We use baselines from five categories:
\begin{packeditemize}
\item {\em Uniform quality:} AWStream~\cite{awstream} tunes the encoding parameters of the underlying codec (resolution, QP, and frame rate), though unlike \name, they use the same configuration for all frames in each time window (on the timescale of minutes)\footnote{We assume that AWStream can obtain the accuracy-delay profile without extra cost, which makes AWStream strictly better.}. (VStore~\cite{vstore} shares a similar idea.)
To show their limitation, we use an ``idealized'' version where the parameters are set such that the size reduction is maximized while its accuracy is almost the same to \name.
\item {\em Server-driven approach:} DDS~\cite{dds} and EAAR~\cite{eaar} belong to this type and they share the idea of encoding different regions with different quality levels. By default, we use QP = (40, 30) as the low quality and high quality settings.\footnote{Instead of letting EAAR predict the region proposal on new incoming frames through tracking, we directly let EAAR obtain the new region proposal, which makes EAAR strictly better.}
\item {\em Frame filtering:} We choose Reducto~\cite{reducto}, one of the most recent proposals along this line. We use the implementation from~\cite{reducto-codebase}.
\item {\em Autoencoder}: We pick a pre-trained autoencoder~\cite{cae-codebase} (introduced in~\cite{cae}). 
\end{packeditemize}
We do not include CloudSeg~\cite{cloudseg} in our evaluation, because it augments the server-side DNN by a super-resolution model, which is complementary to the camera-side video encoding schemes above.
We also does not apply \name's 10\% frame sampling (\S\ref{sec:encoding}) to these baselines to reduce their camera-side overhead, because only Reducto and Autoencoder have heavy camera-side overhead, but with 10\% frame sampling, they will simply ignore 90\% of frames and thus have significantly lower accuracy, whereas \name still encodes every frame (though with slightly outdated \AccuracyGradient) and can deliver high accuracy.


\mypara{Metrics}
Following the definitions in \S\ref{subsec:challenge}, we compare different techniques along three key metrics: {\em delay}, {\em inference accuracy}, and {\em camera-side compute cost} (the cost is measured by camera-side encoding delay and overheads).
In particular, we use F1 score as the accuracy metric in object detection, IoU in semantic segmentation, and distance-based accuracy in keypoint detection. These metrics all values in [0,1], with higher values the better.
We calculate the camera-side delay on one Intel Xeon Silver 4110 CPU.
We assume there are 5 video streams sharing a network link with 2.5Mbps bandwidth upload speed (the average upload speed of Sprint LTE connection~\cite{sprint-LTE}) and 100ms latency~\cite{hsrnet}.
We do not include the server-side inference delay, since \name does not put extra compute cost on the server side, and the optimization of server-side delay is not our contribution either.





\vspace{-0.1cm}

\tightsubsection{Better accuracy-delay tradeoffs}
\label{subsec:eval:e2e}

Figure~\ref{fig:eval-overall} compares \name's performance distributions with those of the baselines on the three tasks (on their perspective default full DNNs) and various datasets\footnote{We do not evaluate region-proposal-based approaches like EAAR and DDS on EfficientDet and Yolo since these DNNs have no region proposal (except that we evaluate DDS on Yolo since DDS develops specific heuristics to handle Yolo).}.
We can see that \name outperforms the baselines: 
in terms of delay, \name has 10-43\% smaller encoding delay than the best baselines with comparable accuracy. 
Vigil has lower streaming delay than \name, but it has low accuracy (many small objects are missed).
\name is also 0.5-2\% more accurate when compared to the non-server-driven baselines with lower streaming delay. 
Though some server-driven techniques have higher accuracy than \name on region-proposal-based DNNs like FasterRCNN, they are not applicable to DNNs that lack explicit region proposals like Yolo and EfficientDet.

We also evaluate \name's performance on semantic segmentation with FCN~\cite{fcn-resnet101} as the final DNN. We find that \name has 20\% higher accuracy than Reducto with lower streaming delay, or 5\% lower streaming delay while maintaining higher accuracy than AWStream. 
This improvement may seem marginal, but the actual improvement of \name will be higher, since AWStream streams the highest-quality video whenever bandwidth permit to the server to identify the best video encoding decision, which can incur a high delay.



\mypara{\Edit Autoencoder}
We also compare \name to the autoencoder~\cite{cae} for object detection on the highway dashcam videos. 
\name achieves an accuracy of 85\%, but the autoencoder only achieves 62\%. 
Moreover, the encoded frame size of \name is about 7KB, much less than that of autoencoder (240KB~\cite{cae-codebase} per frame).
As a result, the streaming delay of autoencoder is over 38 seconds.
Thus, \name has a much better accuracy-delay tradeoff.








\begin{figure}
    \centering
    \subfloat[][\centering{Delay breakdown on Intel Xeon Silver 4100}]
    {
        \includegraphics[width=0.85\columnwidth]{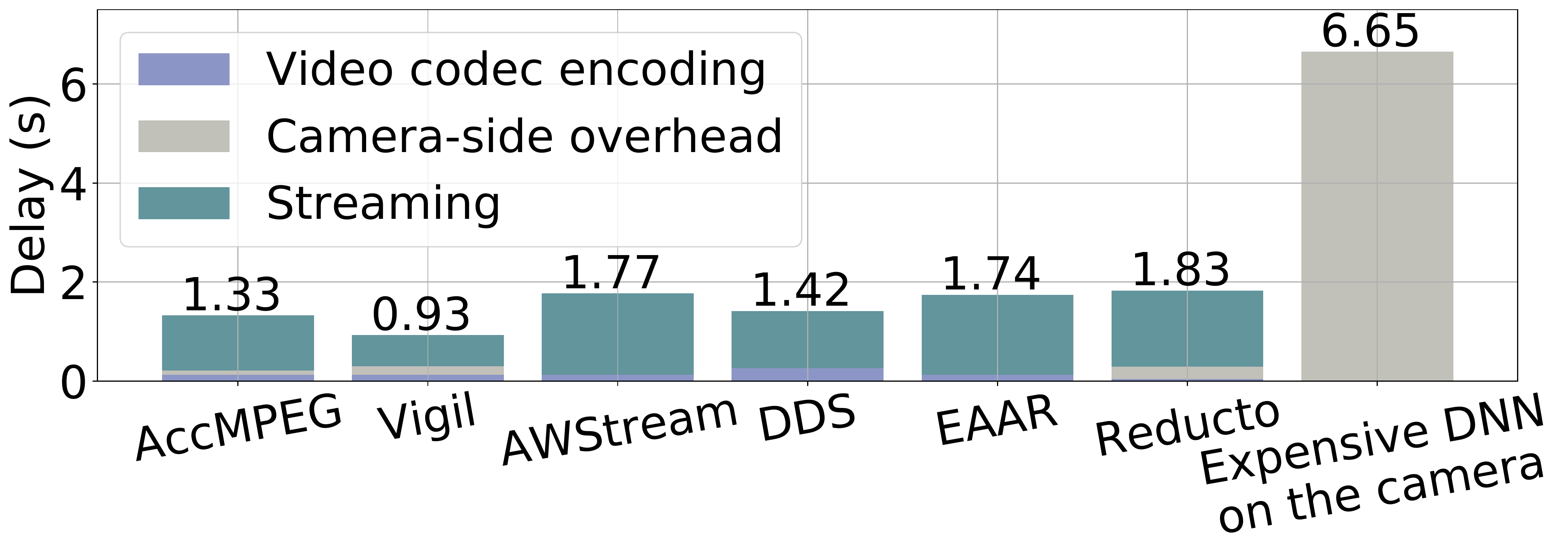}
        \label{subfig:1}
    }
    \vspace{-0.2cm}
    
    \subfloat[][\centering{Delay breakdown on Jetson Nano}]
    {
        \includegraphics[width=0.7\columnwidth]{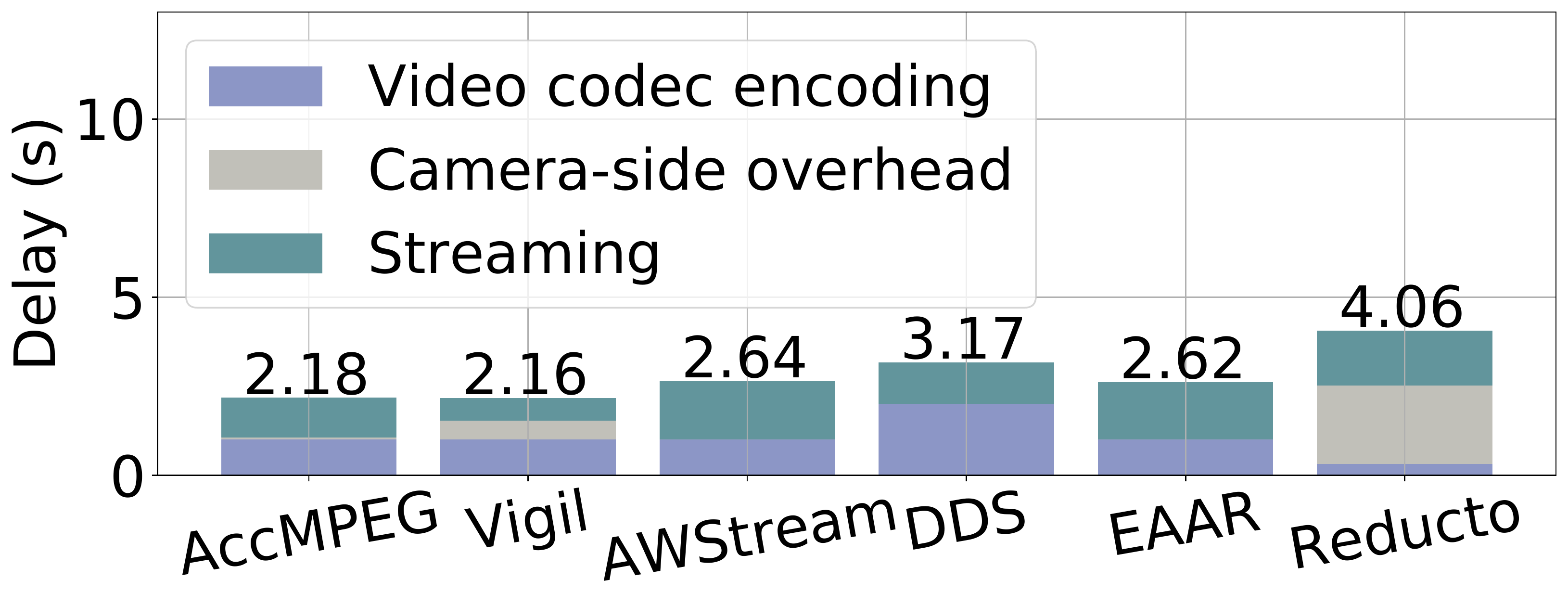}
        \label{subfig:1}
    }
    \tightcaption{Delay breakdowns of \name and baselines. \name achieves minimum streaming delay and has marginally higher encoding delay than codec encoding (used in AWStream, EAAR).}
    \label{fig:eval-delay-overall}
\end{figure}

\tightsubsection{Encoding and streaming delays}
\label{subsec:delay}


\mypara{Delay breakdown}
Figure~\ref{fig:eval-delay-overall} shows the video codec encoding delay, camera-side extra compute delay, and the streaming delay of \name and those of the baselines based on the settings of Figure~\ref{subfig:e} (other settings have similar delay comparisons). 
We can see that \name has the lowest end-to-end delay on both camera-side hardware settings compared to all baselines except Vigi (whose accuracy is much lower than \name in Figure~\ref{fig:eval-overall}).

\mypara{Camera-side compute cost}
We then zoom in on the camera-side compute cost, which consists of encoding delay and the camera-side overhead delay.
Figure~\ref{fig:eval-delay-overall} shows that \name's camera-side \AccuracyGradientModel is cheaper than H264-based video encoding on the CPU, and is 20x cheaper than encoding on Jetson Nano.
Moreover, \name's camera-side compute cost is lower than existing camera-side heuristics, such as Vigil and Reducto.
Compared to Vigil, \name has lower compute cost, because it only runs the camera-side \AccuracyGradientModel inference once every 10 frames, whereas Vigil performs camera-side inference on every frame. 
Compared to Reducto, \name does have higher encoding delay (since Reducto discards some frames and only encodes the remaining ones), but Reducto runs expensive camera-side logic on every frame\footnote{Reducto performs Harris feature extraction, which contains several convolution filters and per-pixel eigen value decomposition and contributes 70\% of the camera-side ovehead.} and thus has a much higher camera-side overhead than \name.

As a reference point, we also test the camera-side overhead of running the expensive DNN on the camera: the camera-side delay is almost 7 seconds on CPU, and the expensive DNN cannot fit into the GPU memory of Jetson Nano.

\begin{figure}[t]
    \centering
    \includegraphics[width=0.7\columnwidth]{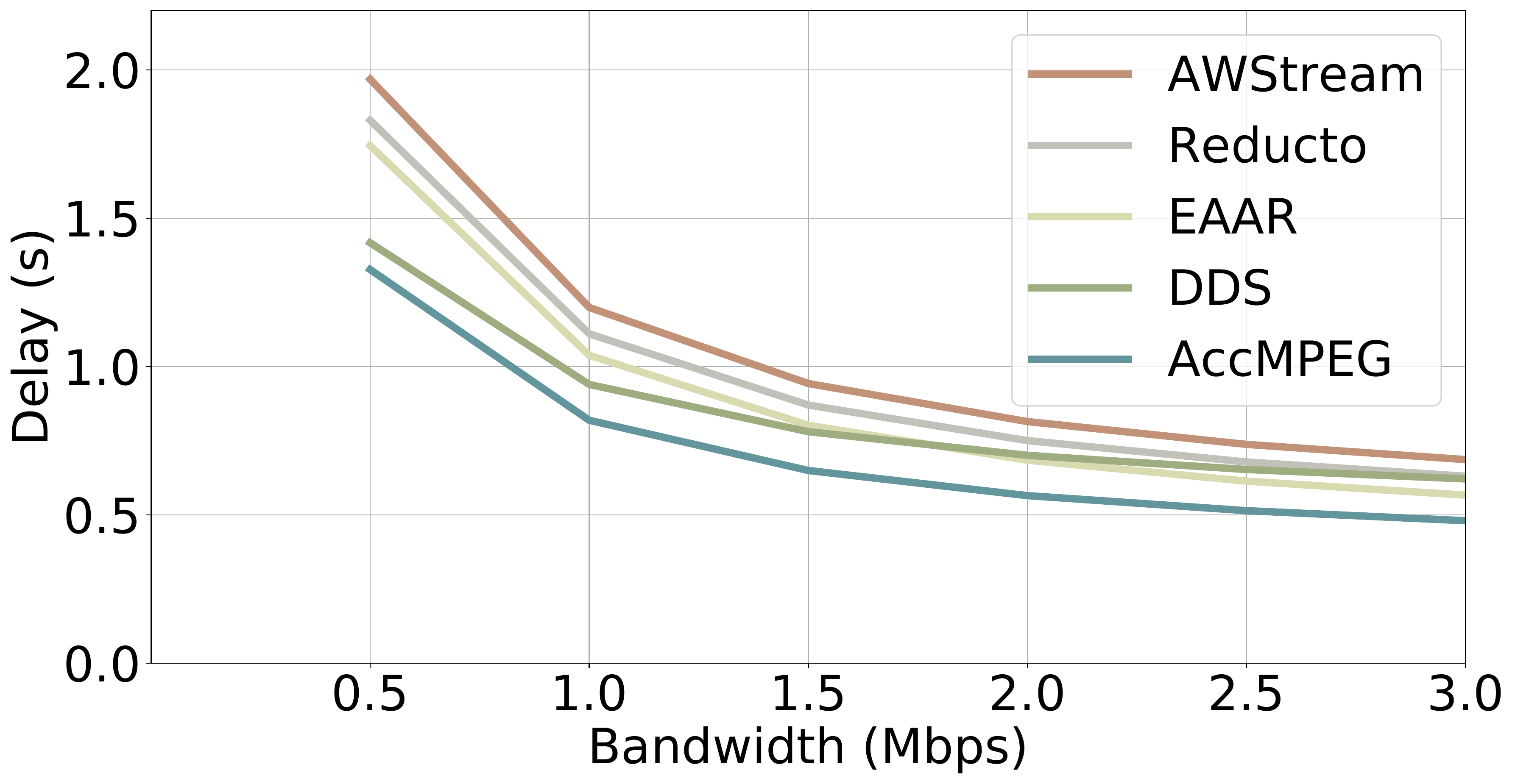}
    \vspace{-0.1cm}
    \tightcaption{The delays of \name and baselines under varying network bandwidth.
    }
    \label{fig:eval-delay-bw}
\end{figure}

\mypara{Delay vs. bandwidth}
Next, we benchmark the impact of network bandwidth on the video-analytics delay, we calculate the delay of \name and various baselines (except for Vigil, which has accuracy lower than 80\% for most of the cases) under increasing network bandwidth.
From Figure~\ref{fig:eval-delay-bw}, we see that \name consistently achieves the lowest delay under different network bandwidth, though with more gains under low bandwidth.




\mypara{Delay optimizations of \name}
\name uses two techniques to speed up its \AccuracyGradient-based encoding: 
(1) using region-of-interest encoding to encode the video (rather than encoding video twice as in DDS~\cite{dds}), and
(2) running the \AccuracyGradientModel model once per 10 frames.
Figure~\ref{fig:eval-cost-reduction} shows their incremental reductions on \name's camera-side delay. 
The figure breaks down the encoding delay of \name into \AccuracyGradient prediction (\AccuracyGradientModel) and the actual codec encoding, and as a reference point, it also shows the encoding delay of H.264, DDS and VP9\footnote{We use the real-time encoding option of VP9.}.
As \name uses the \AccuracyGradientModel (a shallow DNN) for its accuracy gradient model, the delay of accuracy gradient prediction is much smaller than prior work such as DDS which needs to actually run the final DNN. 
That said, it is sizable compared with the encoding delay.
\name further reduces the delay by running \AccuracyGradientModel on one frame every 10 frames, which allows \name to encode frames at 30fps on one Intel Xeon Silver 4100 CPU.




\tightsubsection{Fast \AccuracyGradientModel training and reusing}
\label{subsec:eval:customization}


\mypara{Efficacy of reusing \AccuracyGradientModel}
From Figure~\ref{fig:pretrain}, we see that in object detection, the \AccuracyGradientModel trained on FasterRCNN also provides performance benefit on YoLo and EfficientDet across two different datasets.
Similarly, in keypoint detection, the \AccuracyGradientModel trained on KeypointRCNN-ResNet101 also generalizes to KeypointRCNN-ResNet50 on the surfing dataset.
This demonstrates that \name can generalize to different vision models and provide better accuracy-delay trade-off, as long as the models are trained on the same dataset (as explained in \S\ref{sec:training}).


\begin{table}[t]
\vspace{0.1cm}
\footnotesize{
    \centering
    \begin{tabular}{p{5.cm}|p{1.7cm}}
    \hline
    \hline
    Basic training pipeline (Figure~\ref{fig:training}(a)) & 453 minutes  \\
    \hline
    After decoupling final DNN from training (Figure~\ref{fig:training}(b)) & 74.0 minutes  \\
    \hline
    After 10$\times$ training data downsampling & 7.40 minutes  \\
    \hline
    \end{tabular}
    \tightcaption{The training time of \name on 8 GPUs.
    }
    \vspace{-0.1cm}
    \label{tab:eval-training}
}
\end{table}

\mypara{Fast training}
To benchmark the training speed, we train \AccuracyGradientModel for FasterRCNN~\cite{faster-rcnn} on 10x-downsampled COCO dataset~\cite{coco}. The training takes less than 8 minutes in total on 8 RTX 2080 Super GPU. 
From Table~\ref{tab:eval-training}, we see that downsampling and the \AccuracyGradient abstraction reduces the overall training time by 60x.

\tightsection{\Edit Discussion on generalization of \name}
{
\Edit




While \name improves performance in most cases, \name does not generalize to all video content and may have marginal or negative improvement for some video content.

First, our camera-side cheap quality-selection model may fail when the content of test videos is out-of-distribution (\eg when the videos contain objects of a new class that did not appear in the training set of \AccuracyGradientModel). 
That said, in Figure~\ref{fig:eval-overall}, we empirically show that \name performs reasonably well on YouTube videos, when \AccuracyGradientModel is trained on images sampled from the large training set of the server-side final DNN. 

Second, \name reduces the overhead of \AccuracyGradientModel by using a cheap architecture and running it on sampled frames (\S\ref{sec:encoding}), but these cost optimization techniques may make it fail on certain types of video content. 
As \AccuracyGradientModel is run every 
$k=10$ frames, it will have low accuracy if the objects are moving very quickly (\eg monitoring camera feeds from a race car).
Moreover, the architecture of \AccuracyGradientModel---MobileNet-SSD~\cite{mobilenetv2}---works well on medium to large sized objects, and it can perform poorly with tiny objects (such as distant vehicles in drone videos).



}






\tightsection{Related work}
\label{sec:related}


\mypara{Video analytics pipelines}
There are many proposals to balance video analytics accuracy and its costs, including computing cost (\eg~\cite{videostorm,chameleon,vstore,filterforward,deepcache}) as well as compression efficiency (\eg~\cite{dds,awstream,eaar,vigil}). 
Besides those elaborated elsewhere in the paper, 
other techniques also try to discard unimportant frames~\cite{fvc,glimpse,frugal,filterforward,focus} or downsize the quality/framerate of an entire video segment~\cite{vstore,chameleon,videostorm,haris2018task}, offload inference of RoI bounding boxes~\cite{elf} to remote servers, and raise bitrate in regions found by feeding DNN through the final DNN~\cite{galteri2018video,choi2018high}. 
Again, \name differs in that it introduces a cheap DNN-aware module to perform macroblock-level (rather than object-based) quality optimization and can quickly customize for any given final DNN.




\mypara{Vision feature encoding}
Other video encoders extract vision feature maps from the video and then compress the features (\eg~\cite{duan2020video,xia2020emerging,cracking,neurosurgeon,matsubara2019distilled}), with some efforts to standardize this approach~\cite{gao2021recent,VCM,CDVA}.
Some also optimize for both vision accuracies and human visual quality (\eg~\cite{hu2020towards}).
These video codecs explore a different design point than \name:
(1) they assume that all video analytics DNNs share the same feature extractor (instead, \name treats each final DNN as just a blackbox); 
(2) they redesign both the encoder and the decoder (instead, \name run on any standard video codec); and 
(3) Their target vision tasks (\eg classification or action recognition) have more error tolerance when compressing feature maps (instead, \name handles more expensive tasks, like object detection, where any distortion on the feature maps matters). 

\mypara{Deep learning-based video compression}
Some parallel efforts also replace the video codec by autoencoders (\eg~\cite{lu2019dvc,habibian2019video,agustsson2020scale,rippel2019learned,wu2018video}).
In a similar spirit, recent work trains differentiable video encoders to improve inference accuracy on the decompressed videos (\eg~\cite{chamain2021end}). 
These DNN-based autoencoders do not directly apply, since these autoencoders
are orders of magnitude more expensive than the standard video codes (used in \name): the fastest autoencoder runs at similar speed on GPU as H264 on CPU~\cite{rippel2019learned}.

\mypara{Adapting spatial scales in computer vision}
The computer vision community also uses adaptive image sizing or partitioning to improve inference accuracy; \eg feature pyramid networks (FPN)~\cite{lin2017feature} and BiFPN~\cite{tan2020efficientdet} extract feature maps from multiple resolutions to detect small objects.
Others use attention mechanisms to focus computation on regions with potential objects~\cite{wang2017residual,ozge2019power,ruuvzivcka2018fast,fan2019shifting}.
While \name shares similar insights,
it optimizes the video compression efficiency, rather than computation complexity. 

\tightsection{Conclusion}
In this work, we present \name, a new video codec for video analytics that improves the tradeoffs between inference accuracy and compression efficiency for a variety of computer vision tasks.
It does so by treating any vision DNN as a {\em differentiable black box} and infers the {\em accuracy gradients} to identify where in the frame the DNN's inference result is highly sensitive to the encoding quality level and thus needs to be encoded with high quality.
Our evaluation of \name over three vision tasks shows that compared with the state-of-the-art baselines, \name reduces upto 43\% of the delay while increasing accuracy by upto 3\% at the same time. Moreover, \name's camera-side overhead is almost the same as those of the traditional codecs.

\tightsection{Acknowledgments}

The paper is partly supported by NSF CNS-2146496, CNS-2131826, CNS-1901466, and a  Google Faculty Research Award. Thank Yuhan Liu for her proofreading. Hope Xiulan Xu would smile if she could see this work.

\bibliographystyle{mlsys2022}
\bibliography{reference}






\appendix

\appendix
\section{Artifact Appendix}
\label{appendix:AE}
\subsection{Abstract}

This artifact includes the implementation of \name, a static ffmpeg binary as well as a modified ffmpeg source code.

\subsection{Artifact check-list (meta-information)}

{\small
\begin{itemize}
  \item {\bf Algorithm: yes}
  \item {\bf Binary: ffmpeg} (we use ffmpeg 3.4.8, but we also reproduced our results under ffmpeg 5.0)
  \item {\bf Model: MobileNet-SSD} (we provide the model weight in our github repository)
  \item {\bf Data set: downloaded from youtube}
  \item {\bf Run-time environment: Ubuntu 18.04 with CUDA available. Root access not required.}
  \item {\bf Hardware: NVIDIA GPU}
  \item {\bf Metrics: delay and accuracy}
  \item {\bf Output: delay-accuracy.jpg}
  \item {\bf How much disk space required (approximately)?: 25GB}
  \item {\bf How much time is needed to prepare workflow (approximately)?: 2 hours}
  \item {\bf How much time is needed to complete experiments (approximately)?: 3 hours}
  \item {\bf Publicly available?: Yes}
  \item {\bf Code licenses (if publicly available)?: Apache-2.0}
  \item {\bf Data licenses (if publicly available)?: none}
  \item {\bf Workflow framework used?: no}
  \item {\bf Archived (provide DOI)?: \name (\href{https://doi.org/10.5281/zenodo.6047842}{https://doi.org/10.5281\\/zenodo.6047842}), ffmpeg 5.0 (\href{https://doi.org/10.5281/zenodo.6048006}{https://doi.org/10.5281/zenodo.\\6048006}), modified ffmpeg source code (\href{https://doi.org/10.5281/zenodo.6051544}{https://doi.org/10.5281/\\zenodo.6051544})}
\end{itemize}}

\subsection{Description}

\subsubsection{How delivered}

\begin{packeditemize}
    \item 
    Implementation of \name: \href{https://github.com/KuntaiDu/AccMPEG}{\name github repository}, MLSys branch
    \item
    Modified ffmpeg source code:
    \href{https://github.com/Alex-q-z/myh264}{Modified ffmpeg github repository}, AccMPEG branch
    \item 
    Static ffmpeg binary:
    \href{https://johnvansickle.com/ffmpeg/}{Johnvansickle ffmpeg}
\end{packeditemize}

\subsubsection{Hardware dependencies}

NVIDIA GPU.

\subsubsection{Software dependencies}

GCC, NVIDIA CUDA driver and conda.

\subsubsection{Data sets}

We download all of our videos from youtube. Please check \href{https://docs.google.com/spreadsheets/d/15sJ1yt86OuNqVx-WjFmisRa-UkPzoHGX8mfzE3Sog2k/edit#gid=0}{this google spreadsheet} for the details on how we collected the dataset.

\subsection{Installation}

Please refer to \href{https://github.com/KuntaiDu/AccMPEG/blob/MLSys/INSTALL.md}{INSTALL.md} to install our code.

\subsection{Experiment workflow}

Please refer to \href{https://github.com/KuntaiDu/AccMPEG/blob/MLSys/README.md}{README.md} for the experiment workflow of our code.

\subsection{Evaluation and expected result}

Please refer to \href{https://github.com/KuntaiDu/AccMPEG/blob/MLSys/README.md}{README.md} for the evaluation and expected result of our code.

\section{Optimal quality assignment analysis}
\label{appendix:accuracy-gradient}

We formalize the spatial quality assignments in this section and derive the near-optimal solution through \AccuracyGradient.

\subsection{Formalize quality assignment}
To make the discussion more concrete, we split each $\OriginWidth\cdot\OriginHeight$ frame into $\GridWidth\cdot\GridHeight$ grids of 16x16 blocks ($\OriginWidth=16\GridWidth,\OriginHeight=16\GridHeight$) and assign each block either a high quality or a low quality. 
We now consider this problem: what is the best quality assignment for these 16x16 blocks that maximizes the accuracy subject to no more than $\MaxHighQualityFrac$ blocks encoded in high quality.
Formally, it searches for a binary mask $\Mask$ of size $\GridWidth\cdot\GridHeight$ ($\Mask_{x,y}=1$ means block $x,y$ is in high quality), such that 

{
\vspace{-0.2cm}
\footnotesize
\begin{align}
\textrm{max} &~~~~\SimFunc\left(\DNN(\Mask\times\HighImage+(\AllOnes-\Mask)\times\LowImage),\DNN(\HighImage)\right) \label{eq:idealized-objective}\\
\textrm{s.t.} & ~~~~ \Vert\Mask\Vert \leq \MaxHighQualityFrac
\end{align}
}
where $\HighImage$ and $\LowImage$ are the high-quality encoding and the low-quality encoding of each frame\footnote{The dimension of a frame is the same for different QP values, so $\HighImage$ and $\LowImage$ have the same dimension.}, $\DNN:\mathbb{I}\mapsto\mathbb{O}$ returns the DNN inference result ($\mathbb{I}$ and $\mathbb{O}$ are the spaces of input frames and DNN output), and $\SimFunc:\mathbb{O}\times\mathbb{O}\mapsto\mathbb{R}$ returns the accuracy of $\DNN$'s output on a compressed frame by comparing its similarity with $\DNN$'s output on the high quality image $\DNN(\HighImage)$.

This formulation involves two simplifying assumptions: the 16x16 blocks may be suboptimal boundaries between quality levels, and it restricts the encoding to only two quality levels.
That being said, we believe that analyzing this formulation is still valuable for two reasons.
First, the block granularity of 16x16 is on par with the block sizes employed in H.264 and H.265, which means more fine-grained blocks will not have much impact on the encoded video size. 
Second, the use of two quality levels does subsume many recent solutions (\eg~\cite{vigil, dds, awstream, glimpse, reducto}) which use two or fewer quality levels.

\subsection{Deriving near-optimal solution through \AccuracyGradient}

In this section, we ``derive'' the near-optimal quality assignment.
We use $\Mask$ to represent the quality assignment over each macroblock $\Block$. $\Mask_\Block=1$ when $\Block$ is encoded in high quality, and $\Mask_\Block=0$ when encoded in low quality.
We then encode the image $\Image$ according to the quality assignment $\Mask$. We assume that  $\Image=\Mask\times\HighImage+(\AllOnes-\Mask)\times\LowImage$, where $\HighImage$ is the high quality image, $\LowImage$ is the low quality image and $\times$ means element-wise multiplication.

Our goal is to maximize the accuracy $\SimFunc$ of our approach against the ground truth $\DNN(\Image)$ ($\DNN$ represents the final video analytic DNN). In other words, we are maximizing $\SimFunc(\DNN(\Image), \DNN(\HighImage)$.
We notice that $\textrm{max}  ~~ \SimFunc(\DNN(\Image), \DNN(\HighImage))$
is equivalent to the following (note that the first term of Eq~\ref{eq:objective} is a constant):
\begin{figure}[t]
    \centering
    \subfloat[\centering{Traditional loss (y-axis) used to train segmentation models.}]
    {
        \includegraphics[width=0.48\columnwidth]{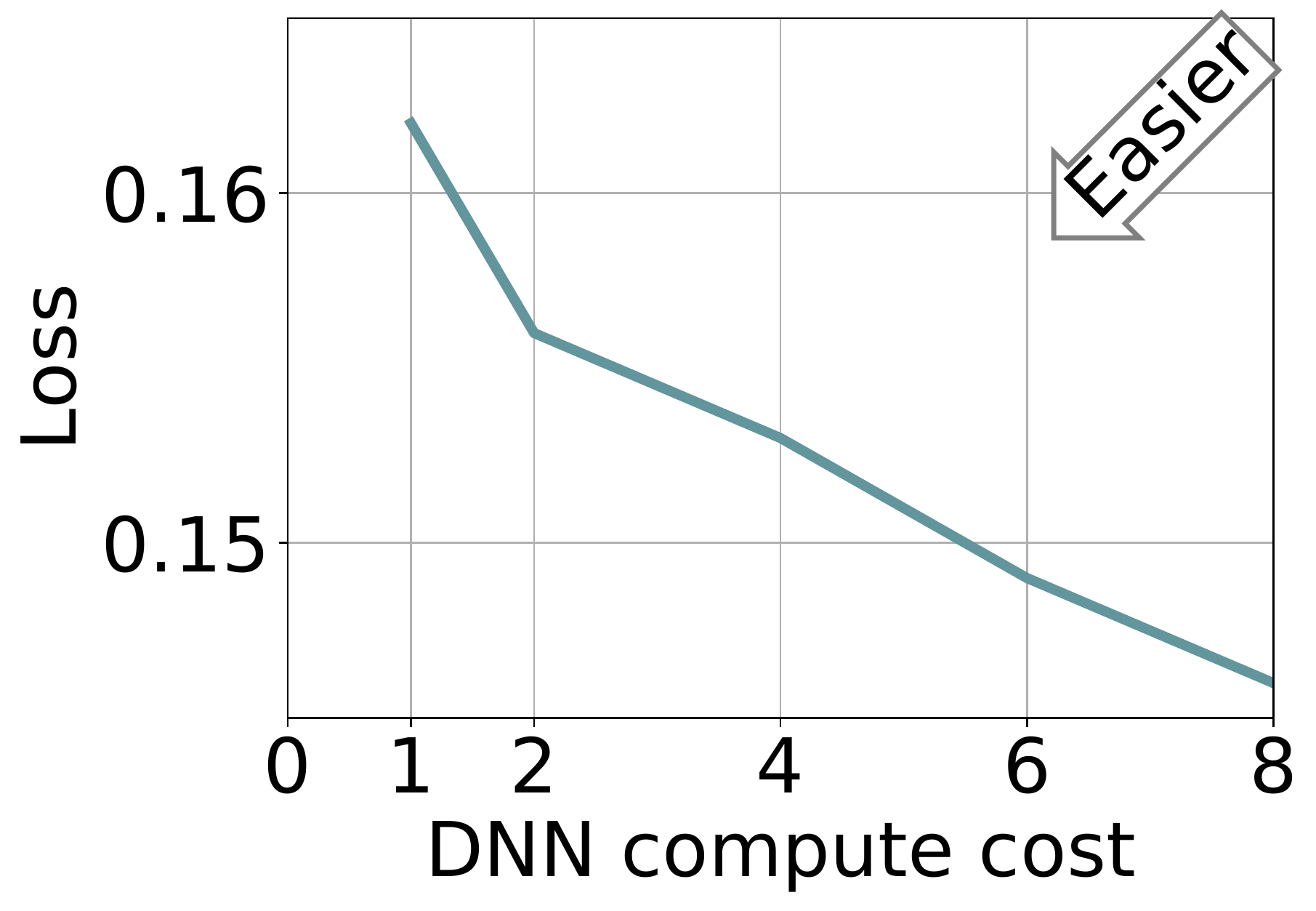}
        \label{subfig:segmentation_training_loss}
    }
    \subfloat[\centering{The loss function (y-axis) used to train \AccuracyGradientModel.
    }]
    {
        \includegraphics[width=0.48\columnwidth]{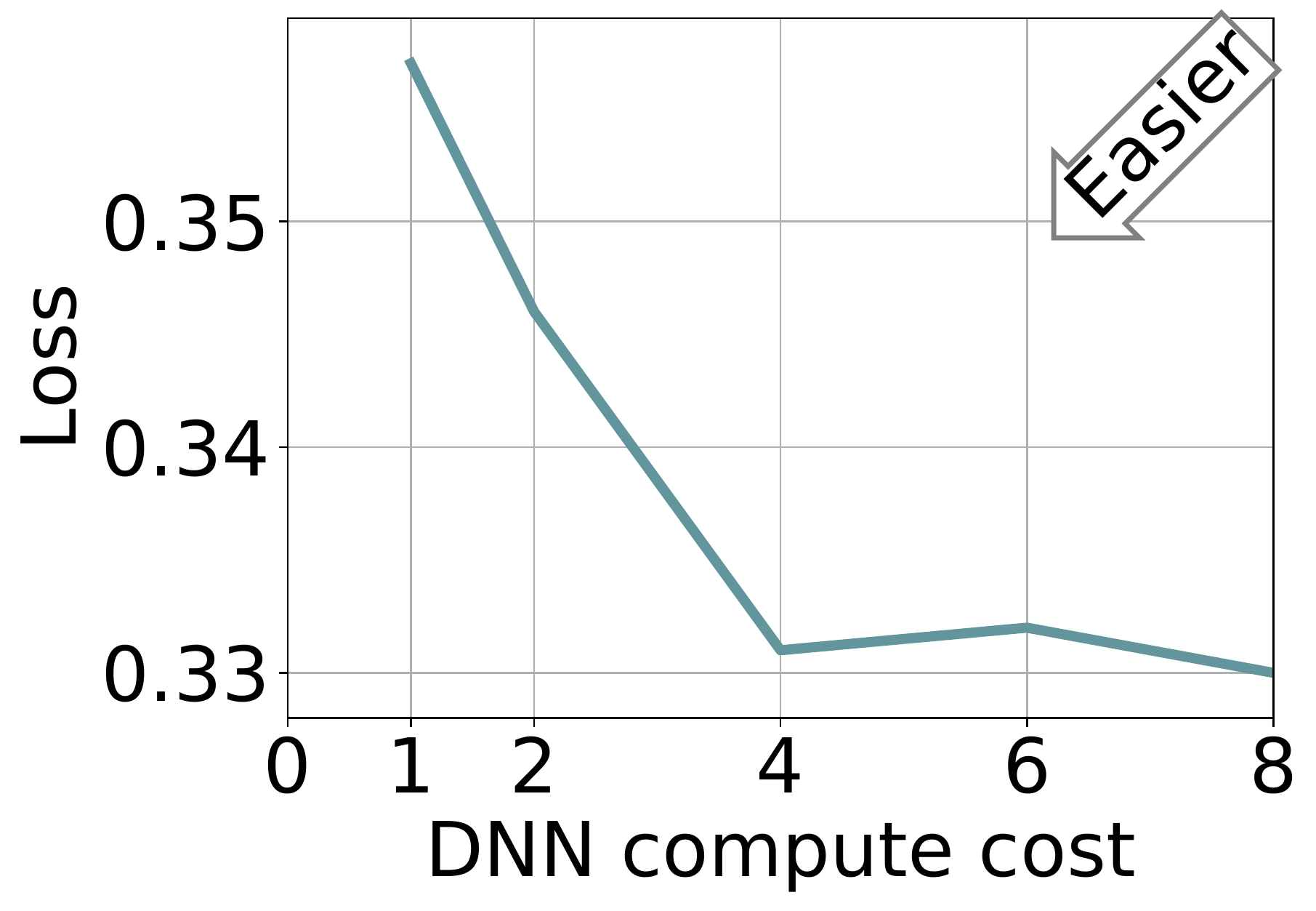}
        \label{subfig:accmpeg_training_loss}
    }
    \tightcaption{
    Comparing the training curve of a traditional segmentation model and that of \AccuracyGradientModel: with more DNN layers (higher DNN compute cost), the traditional segmentation loss drops slowly, whereas the loss function of \AccuracyGradientModel (\ie low-dimensional binary segmentation with a higher tolerance to false positives) drops very quickly, which indirectly suggests that a cheap model might suffice to train an accurate enough \AccuracyGradientModel.
   }
    \label{fig:eval-loss}
\end{figure}

{\footnotesize
\vspace{-0.2cm}
\begin{align}
\textrm{min} ~~~~ & \SimFunc(\DNN(\AllOnes\times\HighImage), \DNN(\HighImage))-\SimFunc(\DNN(\Image), \DNN(\HighImage)) \label{eq:objective}\\
= & \bigg\langle\frac{\partial \SimFunc(\DNN(\Image'), \DNN(\HighImage))}{\partial \Image'},(\AllOnes\times\HighImage - \Image)\bigg \rangle_{\textrm{F}} \label{eq:lagrange}\\
\approx & \bigg\langle\frac{\partial \SimFunc(\DNN(\LowImage), \DNN(\HighImage))}{\partial \LowImage},(\AllOnes\times\HighImage - \Image)\bigg\rangle_{\textrm{F}} \label{eq:relax} \\
= & \bigg\langle\frac{\partial \SimFunc(\DNN(\LowImage), \DNN(\HighImage))}{\partial \LowImage},(\AllOnes-\Mask)\times(\HighImage-\LowImage)\bigg\rangle_{\textrm{F}}, \label{eq:temp}
\end{align}
}
where $\langle\cdot,\cdot\rangle_{\textrm{F}}$ is the Frobenius inner product and $\AllOnes$ is the matrix with all of its elements be 1. 
Eq (\ref{eq:lagrange}) uses Lagrange's Mean Value Theorem \footnote{As the accuracy metric is typically not differentiable, we use the training loss function of the neural network as the differentiable approximation of the accuracy.}, where $\Image'$ lies between $\HighImage$ and $\Image$. In Equation (\ref{eq:temp}), since the value of $\Mask$ is identical inside each block, each block $\Block$ contributes the following value to Eq (\ref{eq:temp}):

{\footnotesize
\vspace{-0.2cm}
\begin{align}
& (1-\Mask_\Block)\sum_{\Pixel\in \Block} \bigg( \frac{\partial \SimFunc(\DNN(\LowImage), \DNN(\HighImage))}{\partial \LowImage}\bigg|_\Pixel \cdot (\HighImage_\Pixel-\LowImage_\Pixel) \bigg) \nonumber\\
\leq & (1-\Mask_\Block) \cdot \sum_{\Pixel\in \Block} \bigg\Vert\frac{\partial \SimFunc(\DNN(\LowImage), \DNN(\HighImage))}{\partial \LowImage}\bigg|_\Pixel\bigg\Vert_1 \cdot \Vert \HighImage_\Pixel-\LowImage_\Pixel \Vert_1 \nonumber\\
= &  (1-\Mask_\Block) \cdot AccGrad_\Block, \label{eq:deliver_accuracy_gradient}
\end{align}
}
where $\Pixel$ means the pixel inside the macroblock $\Block$. The equality condition of the inequality above is the sign of the gradient term $\partial \SimFunc$ aligns with the sign of $\HighImage_\Pixel - \LowImage_\Pixel$, which actually is most of the cases since pushing a pixel closer to high quality typically improves accuracy. After the inequality, we transform the original optimization objective (Eq (\ref{eq:objective}) to minimizing the \textit{quality drop} $1-\Mask_\Block$ times the accuracy gradient $\AccuracyGradient_\Block$. Thus, the optimal solution is to give those high accuracy gradient macroblocks a low quality drop, which means encode them in high quality.

\section{Empirical evidence on how false-positive-tolerance reduces the cost of \AccuracyGradientModel}
\label{appendix:false-positive-tolerance}

To empirically support that false positive tolerance (mentioned in \S\ref{subsec:accuracy-gradient}) can reduce the compute demand of segmentation task, we train a series of DNNs with compute power [1,2,4,6,8]x on the same dataset with two different losses: traditional segmentation loss and our training loss (with less penalty on those blocks that wrongfully encoded in high quality). From Figure~\ref{fig:eval-loss}, we see that 4x-compute DNN performs much worse than 8x DNN under traditional segmentation loss (as shown in Figure~\ref{subfig:segmentation_training_loss}) but the performance of 4x-compute DNN and 8x-compute DNN is similar under our training loss (see Figure~\ref{subfig:accmpeg_training_loss}).
This indicates that our training loss is much less compute-hungry than traditional segmentation loss.

\end{document}